\documentclass[%
 reprint,
 amsmath,amssymb,
 aps,
a4paper
]{revtex4-2}
\usepackage{etoolbox}
\usepackage{mlmodern}
\usepackage[utf8]{inputenc}
\inputencoding{utf8}
\usepackage{chemformula}
\usepackage{graphicx}
\usepackage{enumitem}
\usepackage{gensymb}
\usepackage{amsmath}
\usepackage{amsthm}
\usepackage{amssymb}
\usepackage{dcolumn}
\usepackage{color}
\usepackage{comment}
\usepackage{nameref} 
\usepackage{bbm}
\usepackage{hyperref}
\usepackage{multirow}
\usepackage{xr}
\makeatletter

\usepackage[normalem]{ulem}
\newcommand{\bc}[1]{{\color{blue} #1 }}

\newcommand*{\addFileDependency}[1]{
\typeout{(#1)}
\@addtofilelist{#1}
\IfFileExists{#1}{}{\typeout{No file #1.}}
}\makeatother

\newcommand*{\myexternaldocument}[1]{%
\externaldocument{#1}%
\addFileDependency{#1.tex}%
\addFileDependency{#1.aux}%
}

\myexternaldocument{SI/SI}

\begin{document}

\title{Individual and cooperative superexchange enhancement in cuprates}

\author{Tonghuan Jiang}
\affiliation{School of Physics, Peking University, Beijing 100871, P. R. China}

\author{Nikolay A. Bogdanov}
\email{n.bogdanov@fkf.mpg.de}
\affiliation{Max Planck Institute for Solid State Research, Heisenbergstrasse 1, 70569 Stuttgart, Germany}

\author{Ali Alavi}
\email{a.alavi@fkf.mpg.de}
\affiliation{Max Planck Institute for Solid State Research, Heisenbergstrasse 1, 70569 Stuttgart, Germany}
\affiliation{University of Cambridge, Lensfield Road, Cambridge CB2 1EW, United Kingdom}

\author{Ji Chen}
\email{ji.chen@pku.edu.cn}
\affiliation{School of Physics, Peking University, Beijing 100871, P. R. China}
\affiliation{Interdisciplinary Institute of Light-Element Quantum Materials and Research Center for Light-Element Advanced Materials,Peking University, Beijing 100871, P. R. China
}
\affiliation{Frontiers Science Center for Nano-Optoelectronics, Peking University, Beijing 100871, P. R. China}

\date{\today}

\begin{abstract}

%
It is now widely accepted that the antiferromagnetic coupling within high temperature superconductors strongly exhibits a profound correlation with the upper limit of superconducting transition temperature these materials can reach.
Thus, accurately calculating the positive and negative mechanisms that influence magnetic coupling in specific materials is crucial for the exploration of superconductivity at higher temperatures. 
%
%
%
Nevertheless, it is notoriously difficult to establish a complete description of electron correlations employing \textit{ab initio} theories because of the large number of orbitals involved. 
In this study, we tackle the challenge of achieving 
high-level \textit{ab initio} wave function theory calculations, which allow an explicit treatment of electron correlations associated with a large number of high-energy orbitals. 
We elucidate the atomic-shell-wise contributions to the superexchange coupling in the lanthanum cuprate, including individual effects of high-energy orbitals (Cu 4d, 5d, 4f, 5p) and cooperative effects between the core and these high-energy orbitals. 
Specifically, the prominent contributions from Cu 4d, 5d, 4f and 5p give rise to a rich collection of previously unexamined superexchange channels.  
We propose a $p$-$d$-$f$ model to universally account for the contributions of high-energy orbitals at copper sites. 
Our calculations and physical rationalizations offer a more robust theoretical foundation for investigating cuprate-type high-temperature superconductors.

\end{abstract}

\maketitle


%
Cuprate unconventional superconductors have received widespread attention in fields of condensed matter physics
due to their unique role as the first systems displaying superconductivity above liquid nitrogen temperature
\cite{lee_doping_2006,plakida_high-temperature_2010,zhou_high-temperature_2021}.
Despite extensive research and significant progresses in the past four decades, the precise microscopic mechanisms, including material-specific factors, which underlie the superconductivity have not been fully clarified.
The general consensus is that the necessary pairing mechanism in unconventional superconductivity is mediated by strong spin fluctuations in the anti-ferromagnetic correlations of these systems, which provide an effective ``glue" for pairing.  
In this regard, the very large observed superexchange interaction of cuprate materials is highly relevant, and understanding the factors which leads to unusually large superexchange is an important question. 
Ab initio calculations based on wavefunction methods can play a key role in this regard, partly because they are material-specific, and partly because the description of electronic correlations can systematically controlled and analysed at various levels of theory. 
This allows both predictive calculations on real systems,  as well as obtaining insights into competing effects which give rise to the superexchange, which are very difficult to obtain otherwise \cite{Coen_magnetic_2000,Malrieu_magnetic_2014,Bogdanov_NPcuprate_2022,cui_systematic_2022,katukuri_electronic_2020,katukuri_ab_2022,mcmahan_calculated_1988,das_electronic_2009,jarlborg_bands_2011,lane_antiferromagnetic_2018}. 
%
%

%
%
%

\begin{figure*}
    \centering
    \includegraphics[width=16.0cm]{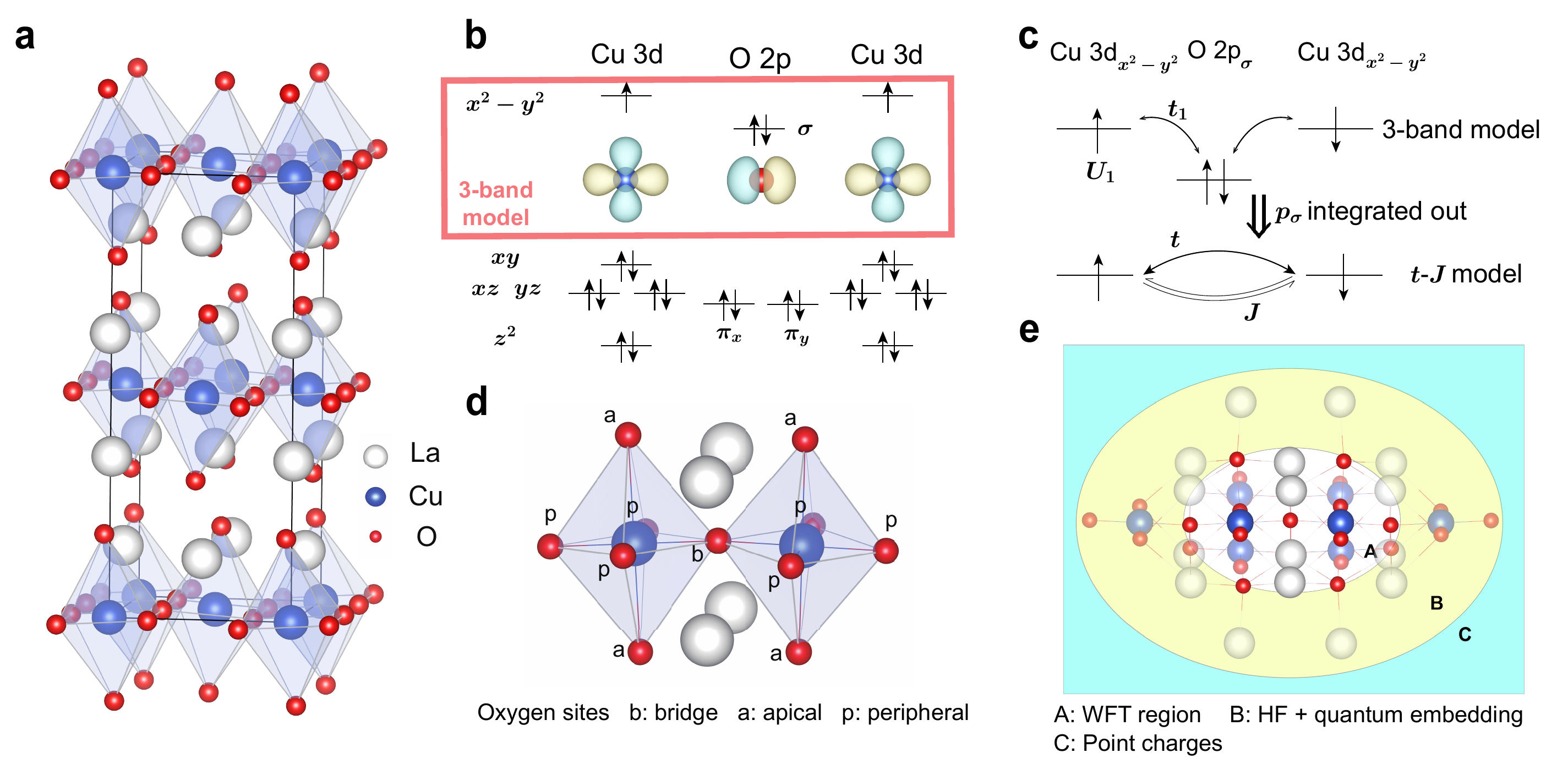}
    \caption{
    Computational and theoretical models.
    (a) Crystal structure of $\mathrm{La_2CuO_4}$. 
    (b) Key aspects of the electronic structure of cuprates. 
    The leading electron configuration of parent compound's ground state is shown, where Cu$^{2+}$ and O$^{2-}$ have 3d$^9$ and O$^{2-}$ 2p$^6$ configurations, respectively. 
    The pink box highlights the orbitals considered in the 3-band model.
    (c) Effective low-energy models derived on key orbitals in cuprates, i.e., Cu 3d$_{x^2-y^2}$ and O 2p$_\sigma$, including the 3-band model and the $t$-$J$ model.
    (d) The $\mathrm{Cu_2O_{11}La_4}$ cluster containing two nearest-neighbor Cu sites and 3 types of oxygen (bridge, apical and peripheral). 
    (e) The three-layer embedding scheme adopted in this work. The correlated WFT calculations are performed on the $\mathrm{Cu_2O_{11}La_4}$ cluster (A, white background), and the environment is split into two layers: the inner quantum projection embedding part (B, yellow shade), and the outer classical point-charge embedding part (C, cyan shade). 
    }
    \label{Fig1_structure}
\end{figure*}

%
Cuprates crystals are formed of copper-oxygen $\mathrm{CuO_2}$ planes and an intercalated ionic bath (Fig. \ref{Fig1_structure}a), which serves mainly as charge reservoirs.
In undoped $\mathrm{CuO_2}$ planes, Cu and O formally have $+2$ and $-2$ valence, respectively.
Each $\mathrm{Cu^{2+}}$ has a $\mathrm{3d^9}$ configuration, with one hole occupying the $\mathrm{3d}_{x^2-y^2}$ orbital to form a spin-1/2 site, 
forming an antiferromagnetic (AFM) ground state (Fig. \ref{Fig1_structure}b)\cite{lee_doping_2006,plakida_high-temperature_2010}.
In the doped case, long-range AFM order breaks down quickly, yet the extra charges from the ion bath lead to Cooper pair formation in $\mathrm{CuO_2}$ planes, in which spin-fluctuations are nevertheless believed to play a key role. \cite{lee_doping_2006,zhou_high-temperature_2021,plakida_high-temperature_2010}. 
Recently Wang et al. proposed an empirical linear dependence between critical temperature ($T_\text{c}$) and AFM coupling in a family of Hg-based cuprates, where a 1 meV enhancement of AFM coupling is accompanied by several Kelvins of $T_\text{c}$ increase \cite{wang_paramagnons_2022}. 
The empirical linear relationship has been further supported by the theoretical work of Qin et al. \cite{qin_intrinsic_2025}, which demonstrated that the maximum superconducting transition temperature ($T_c$) of unconventional superconductors cannot exceed 0.04 to 0.07 times the pairing interaction strength. 
Given the correlation between exchange interaction $J$ and the upper limit of $T_c$, in the pursuit of higher temperature superconducting materials, it is of utmost importance to accurately compute, with meV-precision, the mechanisms that either enhance or impede $J$.

In recent years, \textit{ab initio} wave function theories (WFT) have been developed rapidly, providing new opportunities to accurately tackle complex materials with strong electron correlations \cite{guther_neci_2020,Zhai_Block2_DMRG_2023,sharma_DMRG_2012,Angeli_NEVPT2_2002,Guo_DMRG_SC_NEVPT_2016,Miralles_DDCI_1993,Pulay_CASPT2_2011,Olsen_CASSCF_2011}.
Such calculations are used to obtain quantitatively accurate descriptions, to check the validity of existing models of cuprates, 
%
%
and to provide better theoretical models to fit  experimental measurements
\cite{van_oosten_heisenberg_1996,Munoz_PRL_minCASlowJ_2000,calzado_proposal_2001,hozoi_fermiology_2008,foyevtsova_ab_2014,wagner_effect_2014,Bogdanov_NPcuprate_2022,cui_systematic_2022}.
The calculations on cuprate superconductivity ($T_\text{c}\sim\text{40 K}$) usually require a meV accuracy on their magnetic coupling in their undoped phase, rather than traditional chemical accuracy (1 kcal/mol $\approx$ 500 K) \cite{lee_doping_2006,plakida_high-temperature_2010}.
Effective models of cuprates, such as the 3-band Hubbard model and the $t$-$J$ model, consider effective renormalized interactions within Cu $3d_{x^2-y^2}$ and O $\mathrm{2p_\sigma}$ orbitals (Fig. \ref{Fig1_structure}c) \cite{zhang_effective_1988,lee_doping_2006}. 
Some studies would include Cu 4s and O $\mathrm{2p_{z}}$ orbitals, but recent WFT analyses suggested a more complex picture,
such that simplifications could lead to a severe underestimation of the AFM superexchange, 
highlighting an essential role of both static and dynamic correlations from high-energy bands
\cite{pavarini_band-structure_2001,plakida_high-temperature_2010,van_oosten_heisenberg_1996,Munoz_PRL_minCASlowJ_2000,Calzado_JCP_2000,hozoi_fermiology_2008,Bogdanov_NPcuprate_2022}. 
For instance, a new superexchange channel due to orbital breathing within the Cu d shell can bring $J$ up to one half of the experimental value in $\mathrm{Sr_2CuO_3}$ and two thirds in $\mathrm{La_2CuO_4}$ \cite{Bogdanov_NPcuprate_2022}.
However, a considerable portion of $J$ ($\approx\!35\%$ in $\mathrm{La_2CuO_4}$, and $\approx\!48\%$ in $\mathrm{Sr_2CuO_3}$) remains unexplained, indicating the existence of uncovered channels in superexchange. 
Moreover, it is desirable to scrutinize the real interactions behind the simplified effective models and parameters.

In this work, we devise a computational framework utilizing three-layer quantum embedding and high-level WFT to achieve an accurate \textit{ab initio} calculation of the nearest-neighbor AFM coupling in the prototypical cuprate parent compound, $\mathrm{La_2CuO_4}$.
Specifically, a $\mathrm{Cu_2O_{11}La_4}$ cluster (Fig. \ref{Fig1_structure}d) is selected out of the $\mathrm{La_2CuO_4}$ crystal for correlated WFT computations. 
The embedding scheme is illustrated in Fig. \ref{Fig1_structure}e. 
The $\mathrm{Cu_2O_{11}La_4}$ cluster (A) is surrounded by a 2-layer environment, the quantum projection embedding (B) \cite{Manby_ProjEmb_2012} and classical point charge bath (C) , which describe the short-range and long-range environment interaction, respectively.
A new spin-averaged Hartree-Fock scheme is designed to describe the antiferromagnetic environment within the mean-field level. 
The WFT methods used include complete active space self-consistent field (CASSCF) \cite{Werner_CASSCF_1980}, full-configuration interaction quantum Monte Carlo (FCIQMC) \cite{booth_origFCIQMC_2009,guther_neci_2020}, the density matrix renormalization group (DMRG) \cite{Zhai_Block2_DMRG_2023,sharma_DMRG_2012} and the strongly contracted second-order \textit{n}-electron valence state perturbation theory (SC-NEVPT2) \cite{Angeli_NEVPT2_2002,Guo_DMRG_SC_NEVPT_2016,Sokolov_DMRG_TD_NEVPT2_2017}.
See the \textit{Methods} Section for further computational details.

In order to determine the AFM coupling $J$, we perform a calculation of states of different spin multiplicity within the cluster containing two magnetic centers \cite{de_graaf_magnetic_2016}.
$J$ is calculated by the energy difference between the lowest spin-singlet state and the lowest spin-triplet state. 
This method was successfully applied to compute magnetic couplings in various transition-metal oxides 
\cite{Munoz_PRL_minCASlowJ_2000,calzado_proposal_2001,Bogdanov_Cd2Os2O7_2013,Pizzochero_2020,Bogdanov_NPcuprate_2022}.
The use of WFT methods allowed us to systematically examine the atomic-orbital characters of the AFM coupling. 
This is done by computing $J$ including different sets of high-energy atomic orbitals in the correlated WFT calculations, and comparing their contribution.
We find that the contribution of high-energy orbitals includes two parts, the \textit{individual} effect and the \textit{cooperative} effect, both having prominent impacts on AFM coupling of cuprates. 
Within the high-energy orbitals, Cu 4d, 5d, 4f and 5p were found to contribute the most. 
Based on the \textit{ab initio} results, we establish a theoretical model to describe the previously unexplored superexchange mechanisms in cuprates. 
%

%
%
%

%
%

\section*{Results}\label{Results}
\subsection*{Superexchange enhancement}
\label{subsect:superexchange}

Previous \textit{ab initio} works have already demonstrated the possibility of superexchange enhancement due to electron correlation effects \cite{Munoz_PRL_minCASlowJ_2000,foyevtsova_ab_2014,Bogdanov_NPcuprate_2022}. 
These 
works pointed to the collective effects of large numbers of high-energy orbitals, often referred to as dynamic correlations \cite{Helgaker_dynamic_2000}.
Here, in order to lay a solid foundation for further discussion on orbital contributions, it is necessary to revisit and clarify the concept of superexchange enhancement using different orbital settings to calculate $J$.
%
%
%
To achieve reliable conclusions, we have employed various wave function methods, the results are summarized in Table \ref{J_whole}.

\begin{figure*}
    \centering
    \includegraphics[width=16.0cm]{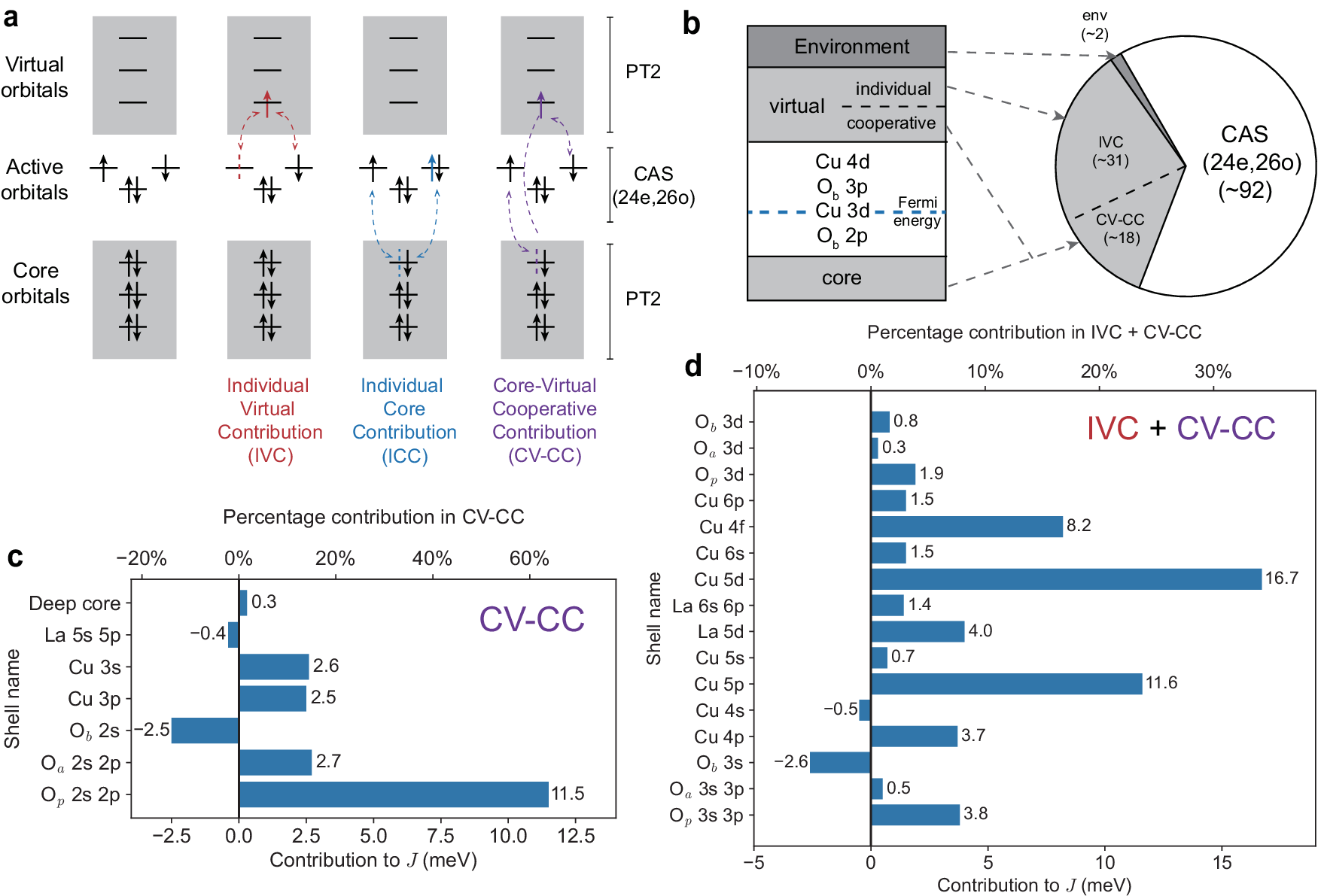}
    \caption{
    Diagrams displaying atomic-shell-wise contributions to superexchange enhancement. 
    (a) Scheme showing different contributions. 
    Orbital contributions beyond  CAS can be divided into 3 parts according to the nature of their electronic configuration:
    ICC, where electrons are excited from core shells to active space; 
    IVC, where electrons are excited from active space to virtual shells; 
    and CV-CC, where both core excitations and virtual excitations occur.
    (b) The weight of different sources of AFM enhancement shown as a pie chart. 
    The multi-reference CAS wave function involving Cu 3d, 4d, and bridging O 2p, 3p adds $\sim$92 meV to AFM coupling. 
    The remaining substantial orbitals contributions are IVC ($\sim$31 meV) and CV-CC ($\sim$18 meV), while ICC ($<$ 1 meV) and 
    environment orbitals beyond Cu$_2$O$_7$ unit ($4\sim$ 2 meV) are negligible. 
    (c) The contribution of different core shells to CV-CC.
    %
    ``Deep core'' denotes atomic shells Cu 1s, Cu 2s, Cu 2p and O 1s. 
    The combined contribution of all core shells equals CV-CC (18.0 meV) and is set as 100\%.
    The summed contribution of all non-intersecting core shells is 16.7 meV. 
    %
    %
    (d) The contribution of different virtual shells to IVC and CV-CC.
    %
    The perturbation space in core orbitals is fixed to the Core-CuO space (see the main text for details). 
    The combined contribution equals IVC + CV-CC (48.8 meV) and is set as 100\%. 
    The summed contribution of all non-intersecting virtual shells is 53.5 meV. 
    The reference space ($\mathcal{B}$) for atomic shell contributions shown in (c) and (d) is discussed in SI Section \ref{SecI_ShellChoice}.
    }
    \label{Fig2_result}
\end{figure*}

\begin{table}[b]
\caption{\label{J_whole}%
AFM coupling $J$ with different WFT calculations. ``PT'' stands for ``perturbation''. The ($m$c,$n$v) in ``PT size'' column means that $m$ core orbitals and $n$ virtual orbitals are correlated in NEVPT2 calculations. Experimental values are taken from spin wave measurements.
The ``ED'' refers to exact diagonalization using Davidson algorithm as the eigenvalue solver. 
%
}
\begin{ruledtabular}
\begin{tabular}{ccccc}
\textrm{CAS size}&
\textrm{CAS solver}&
\textrm{PT space}&
\textrm{PT size}&
$J$ (meV)\\
\colrule
(4e,3o) & ED & None & None & 35.51\\
(24e,26o) & DMRG & None & None & 92.24\\
(24e,26o) & FCIQMC & None & None & 92.9(8)\\
(24e,26o) & DMRG & Full WFT space & (86c,368v) & 143.78 \\
(24e,26o) & DMRG & Frozen env & (86c,180v) & 141.73 \\
(24e,26o) & DMRG & Frozen core & (0c,180v) & 123.74\\
(24e,26o) & DMRG & Frozen virt & (86c,0v) & 91.82\\
& & & Exp. & 120 \cite{Braicovich_SpinWave_2009} \\
& & & & 138 \cite{Coldea_SpinWave_2001}
\end{tabular}
\end{ruledtabular}
\end{table}

First, CASSCF calculations were performed on a small active space, CAS(4e,3o), including only the essential orbitals that correspond the 3-band Hubbard model, i.e. Cu $\mathrm{3d}_{x^2-y^2}$ and bridging O $\mathrm{2p}_\sigma$.
Magnetic coupling obtained from these calculations turns out to be only $J=\mathrm{35.5~meV}$. 
In line with previous studies, such a calculation results in a significant underestimate of superexchange fitted from experimental measurements, which is $\approx\!138$ meV\cite{Bogdanov_NPcuprate_2022,Coldea_SpinWave_2001,Braicovich_SpinWave_2009}.
When the entire Cu 3d, 4d and the bridging O 2p, 3p shells are included in the active space, forming CAS(24e,26o), $J$ is enhanced to 92.2 meV or 92.9 meV using DMRG or FCIQMC solvers respectively. 
These results further support the \textit{orbital breathing} effect, in which the spatial expansion of the effective Cu 3d orbitals due to the correlation with Cu 4d increases the effective $dd$-hopping ($t$) and reduces on-site Coulomb repulsion ($U$), eventually enhancing $J\approx 4t^2/U$ \cite{Bogdanov_NPcuprate_2022}. 
Although the \textit{orbital breathing} captures the leading contribution, 
the calculated $J$ is still 30 - 50 meV away from the reference values
\cite{Coldea_SpinWave_2001,Braicovich_SpinWave_2009}.
%
Once we consider the effect of the whole orbital space using NEVPT2 on top of the CAS(24e,26o) reference wave function,
we find that $J$ increases to 143.8 meV, which is close to the experiment value, manifesting the significant enhancement due to correlations with higher-energy orbitals \cite{Bogdanov_NPcuprate_2022,Coldea_SpinWave_2001,Braicovich_SpinWave_2009}.

%
To further support the validity of embedding treatment, we demonstrate the weak influence of the environment by excluding the empty orbitals of region B from the perturbation space. 
Since all the occupied orbitals in region B have been projected out in the embedding scheme \cite{Manby_ProjEmb_2012}, the remaining environment orbital space consists of only B virtual orbitals.  
The exclusion of B virtual orbitals reduces the number of the correlated virtual orbitals from 368 to 180, yet AFM $J$ is decreased by only 2 meV (143.8 meV to 141.7 meV). 
The result is consistent with the literature where the environment is treated with classical Coulomb potential of point charges and ECPs \cite{Munoz_PRL_minCASlowJ_2000,calzado_proposal_2001,Bogdanov_NPcuprate_2022}. 
%
This confirms that the environment effects on AFM coupling are negligible.
%

Having reproduced all the known correlation effects for superexchange enhancement in cuprates, we are ready to further disentangle the contributions of core and virtual orbitals.
We can conceptually divide the correlation effects of the core and virtual orbitals into three parts (Fig. \ref{Fig2_result}a). 
(i) The part that only involves the virtual orbitals and is unrelated to the core. 
This refers to the contribution resulting from the correlation of active electrons in virtual orbitals, abbreviated as ``individual virtual contribution'' (IVC). 
It is mainly related to the excitation of active electrons to specific virtual orbitals, such as intersite hopping paths or radial diffusion of Cu 3d electrons. 
(ii) The part that only involves the core orbitals and is unrelated to the virtual orbitals. 
This defines the correlation between 
core and active electrons, abbreviated as ``individual core contribution'' (ICC), and is mainly related to hole excitations from active space to core orbitals.
(iii) The part that involves core and virtual orbitals simultaneously, abbreviated as ``core-virtual cooperative contribution'' (CV-CC). 
%
%

Let us define AFM $J$ from CAS(24e,26o) be $J_\text{CAS} = 92.2 \mathrm{~meV}$, and IVC, ICC, and CV-CC to AFM $J$ be $\Delta_\text{IVC}$, $\Delta_\text{ICC}$, and $\Delta_\text{CV-CC}$, respectively. 
Then $J$ correlating different PT orbital sets can be written as  
\begin{equation} 
\begin{aligned}
J_\text{v} & = J_\text{CAS} + \Delta_\text{IVC}, \\
J_\text{c} & = J_\text{CAS} + \Delta_\text{ICC}, \\
J_\text{cv} & = J_\text{CAS} + \Delta_\text{ICC} + \Delta_\text{IVC} + \Delta_\text{CV-CC},
\end{aligned}
\end{equation}
where $J_\text{c}, J_\text{v}, J_\text{cv}$ denote AFM $J$ obtained in three different perturbation 
settings respectively: core space only ($J_\text{c}$), virtual space only ($J_\text{v}$), and core and virtual both ($J_\text{cv}$). 
In other words, CV-CC represents the part of the correlation effect that arises only when both core and virtual correlations are present.

To evaluate sources of different contributions, NEVPT2 calculations on top of the CAS(24e,26o) reference are conducted with different perturbation space selections. 
%
%
We observe that removing virtual orbitals from the perturbation treatment causes $J$ to fall from $J_\text{cv} = 141.7 \mathrm{~meV}$ to $J_\text{c} = 91.8\mathrm{~meV}$, while removing core orbitals
leads to a much smaller reduction, $J_\text{v} = 123.7\mathrm{~meV}$.
In this way, one can see that the correlation effects consist mainly from two parts, IVC ($\sim \mathrm{31~meV}$) and CV-CC ($\sim\mathrm{18~meV}$). 
The individual core contribution, ICC, is in fact negligible ($|\Delta_\text{ICC}|<\mathrm{1~meV}$). 
Hence, core orbitals only have an influence on $J$ when treated together with virtual orbitals, while virtual orbitals lead to an enhancement of $J$ by $\sim\mathrm{31~meV}$ even when they are present alone (Fig. \ref{Fig2_result}b).
In the next two subsections, we discuss IVC and CV-CC effects in detail to uncover individual atomic contributions and discuss possible channels of superexchange.

\subsection*{Atomic-shell-wise correlation effects}\label{subsect:shellwise}

%

First, we discuss how each core and virtual shell contribute to superexchange (Fig. \ref{Fig2_result}c-d). 
The contribution of a set of core orbitals $X$ can be evaluated by comparing the AFM coupling calculation \textit{with} and \textit{without} $X$ included in the correlation treatment. 
\begin{equation}
    \Delta_\mathcal{B}[X] = J[\mathcal{B}] - J[\mathcal{B}\setminus X]
\end{equation}
where the contribution of $X$, $\Delta_\mathcal{B}[X]$, is the difference of $J$ brought by the inclusion of $X$ in the correlated WFT solver, with $\mathcal{B}$ being the corresponding reference space. 
This way, the contribution of $X$ depends on the reference space $\mathcal{B}$. 
%
Therefore, to ensure the validity of comparison, the contributions of different shells are calculated using the same reference space. 
More detailed description of reference space selection can be seen in SI Section I.
After calculating the contribution of several atomic shells $\{X_1,...,X_N\}$, we also check the difference between the summed contribution $\sum_{i=1}^N{\Delta_\mathcal{B}[X_i]}$ and the combined one $\Delta_\mathcal{B}\left[\bigcup_{i=1}^N{X_i}\right]$. 
This difference denotes the non-additive cooperative effects between the selected shells $\{X_1,...,X_N\}$. 
As discussed in the previous section, significant core contributions occur only if both core and virtual shells are correlated, therefore the core contributions listed below belong to CV-CC. 

As can be seen in Fig. \ref{Fig2_result}c,
core orbital contribution to $J$ are mainly concentrated in a set of shallow orbitals, termed ``Core-CuO'', which consists of Cu 3s, 3p, O$_b$ 2s and O$_{(a,p)}$ 2s, 2p.
The subscript in O$_{(b,a,p)}$ denotes the bridging, apical and peripheral oxygens, respectively, as illustrated in Fig. \ref{Fig1_structure}d. 
The rest of core orbitals, i.e. deep-core orbitals (O K, Cu K and L shells), and La orbitals (5s 5p), all exhibit very small contributions to $J$, less than 0.7 meV in total. 
Therefore, Core-CuO encompasses the dominant electron correlation effects within the core shell, and thus is used as the perturbative core space in subsequent calculations on virtual shells.
Among all the core orbital contributions, the peripheral O 2s and 2p shells are the most important ones, contributing $\mathrm{11.5~meV}$. 
We also find that this contribution can be further divided into two parts: $\mathrm{5.8~meV}$ individual effect of 2p, and $\mathrm{5.3~meV}$ cooperative effect between 2s and 2p.
%
Note that the found cooperative effect is a justification for the well known O 2p-Cu 3d correlation.
%
%
%
%
%
%
It is also worth mentioning that the cooperative effects within core shells, measured by the difference between the combined and the summed contributions (Fig. \ref{Fig2_result}c), is negligible (1.3 meV). 
This ensures the validity of discussion about the individual contributions of each atomic orbital shell. 

The contributions of virtual orbitals are investigated the same way, with the only difference being that virtual orbitals are subject to both IVC and CV-CC (Fig. \ref{Fig2_result}d). 
The perturbative core space is set to Core-CuO, such that both the individual and the cooperative virtual orbitals contributions are included. 
Interestingly, unlike the core space, high-energy orbitals in the virtual space do make important contributions. 
For example, the largest contributions of virtual orbitals come from high-energy shells, namely from Cu 5d (16.7 meV), Cu 5p (11.6 meV), and Cu 4f (8.2 meV) orbitals. 
These shells together make up approximately 70\% of the remaining unexplained AFM coupling. 
The effect of Cu 5d is the \textit{orbital breathing}, similar to the effect of Cu 4d. 
%
These high-energy Cu d orbitals are relatively diffuse,
promoting hopping between neighboring sites (increasing $t$) and weakening the on-site repulsion of effective 3d orbitals (reducing $U$). 
Strikingly, our calculations indicate that orbitals with other symmetries, namely Cu 4f and 5p, also contribute to the AFM coupling enhancement.
This will be discussed 
later in the subsection \nameref{subsect:model}. 
%

\subsection*{Cooperative effects between core and virtual orbitals}\label{subsect:cooperative}

\begin{table}[tb]
\caption{\label{J_syner}
The cooperative effects ($\Delta^\text{coop.}[\mathcal{C},X]$) between core and virtual orbitals, evaluated as the difference between calculations with and without core. 
$\mathcal{C}$ stands for ``Core-CuO'' described in the text. 
Details of the reference PT space are presented in Tables \ref{Table1_longlong} and \ref{J_syner_SI} in the SI.
}
%
\begin{ruledtabular}
\begin{tabular}{cccc}
\textrm{virtual shell $X$} &
\textrm{with core} &
\textrm{w/o core}&
\textrm{$\Delta^\text{coop.}[\mathcal{C},X]$} \\
&  $\Delta_{\mathcal{C}}[X]$ &$\Delta[X]$ \\
\colrule
Cu 4f 5p 5d & 34.6 & 35.5 & $-0.8$ \\
O$_b$ 3s O$_p$ 3s 3p Cu 4p & 2.5 & 2.4 & 0.1 \\
La 5d & 4.1 & 0.8 & 3.3 \\
Cu 4s 5s & $-0.1$ & 0.3 & $-0.5$ \\
O$_a$ 3s 3p & 0.5 & 0.2 & 0.3 \\
La 6s 6p & 1.5 & 0.3 & 1.2 \\
Cu 6s 6p & 3.1 & 0.2 & 2.9 \\
O$_b$ 3d & 0.6 & $-8.6$ & 9.1 \\
O$_{(a,p)}$ 3d & 2.1 & 0.5 & 1.6 \\
\colrule
Total & 48.8 & 31.5 & 17.3 \\
\end{tabular}
\end{ruledtabular}
\end{table}

Once we have identified core and virtual shells crucial for exchange, we can further differentiate the individual effects of virtual shells (IVC) from the cooperative effects between core and virtual shells (CV-CC) establishing the basis for an integrated theoretical picture of AFM exchange. 
To achieve this, we calculate the contribution of each virtual shell $X$ with and without core orbitals ($\mathcal{C}$) being correlated, denoted as $\Delta_{\mathcal{B}\cup\mathcal{C}}[X]$ and $\Delta_{\mathcal{B}}[X]$, respectively. 
The cooperative effect between core shell $\mathcal{C}$ and a virtual shell $X$, referred to as $\Delta^\text{coop.}_\mathcal{B}[\mathcal{C},X] = \Delta_{\mathcal{B}\cup\mathcal{C}}[X] - \Delta_{\mathcal{B}}[X]$, is listed in Table \ref{J_syner}. 
One can observe that the virtual shells Cu 4f, 5p, and 5d do not exhibit cooperative effects with the core, despite their large individual effects on $J$.
The shells with significant cooperative effects are oxygen 3d and highly-diffuse orbitals, e.g. Cu 6s 6p, La 5d 6s 6p.
Although perturbation theory usually suggests that orbitals with higher energies play less important role in low-energy physics, our calculations demonstrate that their contributions are not trivial. 

In particular, the contribution of bridging oxygen 3d orbitals (O$_b$ 3d) to AFM coupling shows a peculiar feature.
With core orbitals correlated, the inclusion of O$_b$ 3d orbitals barely changes $J$; however, in the absence of core orbitals, a significant negative contribution due to O$_b$ 3d occurs ($\mathrm{-8.6~meV}$).
A possible explanation may be that the inclusion of O$_b$ 3d alone introduces an extra hopping pathway between nearest-neighbor copper sites; 
because of even parity of O 3d orbitals, the contribution from this two-step hopping has the opposite sign compared to the original $dd$-hopping, thereby weakening AFM coupling. 
When core orbitals are present, the aforementioned effect is suppressed and effectively enhancing the superexchange.
%
%

Computational results of two subsections above are summarized in a schematic diagram presented in Fig. \ref{Fig2_result}b, illustrating the contributions of various atomic orbitals to the AFM coupling. 
This includes a $\sim\!$ 92 meV contribution from CAS(24e,26o), $\sim\!$ 31 meV from the individual contributions of virtual orbitals, $\sim\!$ 18 meV from the cooperative contributions between virtual and core orbitals, and approximately $\sim\!$ 2 meV from the environmental effects. 
Within the individual contributions of virtual orbitals, the prominent correlation effects from Cu high-energy orbitals (4f, 5d and 5p) are found, demonstrating their relevance for the exchange process. 
Among the cooperative contributions between virtual and core orbitals, most significant effects arise from virtual La, high-energy Cu (6s, 6p), and O 3d orbitals. 
%




\subsection*{Extended theoretical models for cuprates}\label{subsect:model}

%
%
%

\begin{figure}
    \centering
    \includegraphics[width=8.5cm]{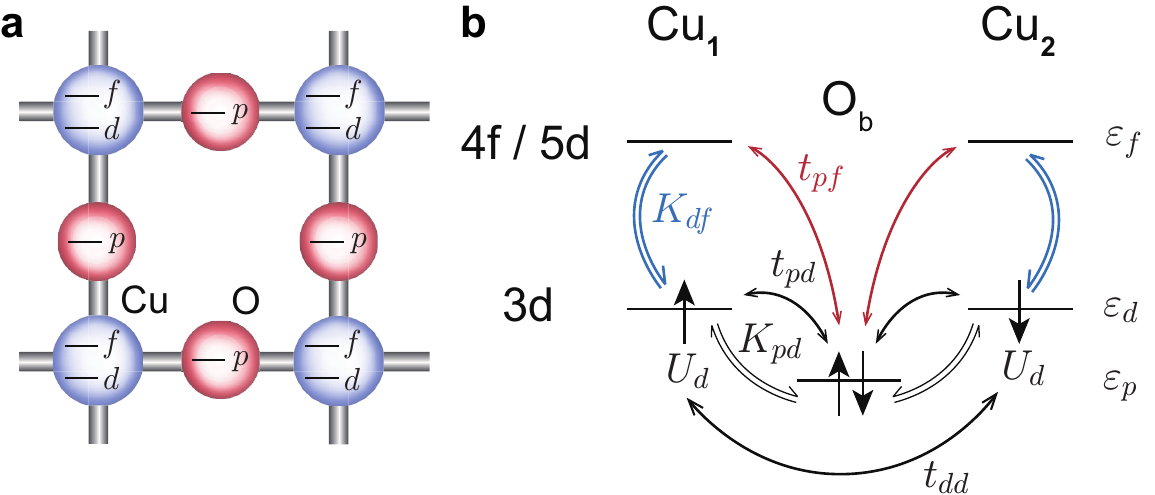}
    \caption{
    Diagram of the $p$-$d$-$f$ model showing Cu 4f and 5p contributions to AFM coupling.
    (a) The periodic version of 4-band $p$-$d$-$f$ model. 
    Each Cu$^{2+}$ ion contains a 3d$_{x^2-y^2}$ orbital (labelled as $d$) and another high-energy orbital which can be either 4f or 5p (labelled as $f$). 
    Each O$^{2-}$ ion has its 2p$_\sigma$ labelled as $p$. 
    (b) The 2-site local version of $p$-$d$-$f$ model. 
    The 5 orbitals in the model is shown in their energy ordering, along with all the inter-orbital integrals in the model. 
    Among these orbitals, the two integrals involving $f$-levels, $K_{df}$ (blue) and $t_{pf}$ (red), are the main source of $f$ participation in AFM coupling enhancement. 
    Other integrals belong to a revision of 3-band model with inter-site exchange integrals, $K_{pd}$, and direct $dd$-hopping, $t_{dd}$, and a more complicated form of $dp$-hopping (SI Section \ref{Sec3_PerturbAnalysis}). 
    }
    \label{Fig3_model}
\end{figure}

%
Following numerical evidence  of prominent contributions of high-energy orbitals, it is instructive to establish an effective model beyond minimal that can faithfully capture elemental exchange mechanisms 
behind the large AFM coupling in cuprates, providing new foundations to study their emergent physics. 
We start with the \textit{orbital breathing} model \cite{Bogdanov_NPcuprate_2022} proposed for Cu 4d, but also valid for Cu 5d, where the on-site exchange integrals serve as a driving force, formulated as 
\begin{equation}
    K = \int{d\mathbf{r}_1d\mathbf{r}_2 \frac{\phi^*_{3d}(\mathbf{r}_1)\phi_{3d}(\mathbf{r}_1)\phi^*_{3d}(\mathbf{r}_2)\phi_{5d}(\mathbf{r}_2)}{|\mathbf{r}_1-\mathbf{r}_2|}}. 
\end{equation}
The correlation of diffuse Cu d orbitals promotes the effective $dd$-hopping and reduces the on-site Coulomb repulsion, and thus enhances $J$. 
However, a substitution of 5d orbital with 4f or 5p in the integral above yields a vanishing $K$ due to the odd parity of 4f or 5p orbitals. 
Therefore, the contribution coming from Cu 4f and 5p cannot be explained within the orbital breathing  mechanism. 
To this end, we propose a further extension of the effective model, termed ``$p$-$d$-$f$ model''.

The 2-site $p$-$d$-$f$ model contains 3 types of orbitals, Cu 3d$_{x^2-y^2}$ ($d$), O 2p$_\sigma$ ($p$), and an additional orbital which can be either Cu 4f or 5p, denoted as $f$ (see Fig. \ref{Fig3_model}). 
The Hamiltonian then includes terms corresponding to the integrals found to be important in \textit{ab initio} calculations. Those are on-site $(U_p, U_d)$ and nearest-neighbor $U_{pd}$ Coulomb repulsion terms, nearest-neighbor hoppings $t_{pd}$, $t_{pf}$ and $t_{dd}$, and exchange integrals $K_{pd}$ and $K_{df}$.
$K_{df}$ (colored in blue in Fig. \ref{Fig3_model}) and $t_{pf}$ (colored in red in Fig. \ref{Fig3_model}) are the only two integrals that involve interaction with high-energy $f$-levels. 
They represent two different superexchange channels that play important role for Cu 4f and 5p contributions, respectively, termed as ``$df$ angular exchange'' and ``$pf$ direct hopping''. 
The integrals are defined as follows. 
\begin{equation}
\begin{aligned}
    K_{df} & = \int{d\mathbf{r}_1d\mathbf{r}_2 \frac{\phi^*_{d}(\mathbf{r}_1)\phi_{f}(\mathbf{r}_1)\phi^*_{d}(\mathbf{r}_2)\phi_{f}(\mathbf{r}_2)}{|\mathbf{r}_1-\mathbf{r}_2|}}  \\
    t_{pf} & = \int{d\mathbf{r}_1d\mathbf{r}_2 \phi^*_{p}(\mathbf{r}_1) \hat{H}_\text{1-body}(\mathbf{r}_1,\mathbf{r}_2)\phi_{f}(\mathbf{r}_2)} 
\end{aligned}
\end{equation}
We can split the 
Hamiltonian for the 2-site $p$-$d$-$f$ model into an unperturbed $\hat{H}_0$ and a perturbation $\hat{H}_1$. 
$\hat{H}_0$ includes orbital energy and Coulomb repulsion terms: 
\begin{equation}
\begin{aligned}
    \hat{H}_0 & = \epsilon_d\sum_{L=1}^2{\hat{n}_{Ld}} + \epsilon_p{\hat{n}_p} + \epsilon_f\sum_{L=1}^2{\hat{n}_{Lf}}  + U_d\sum_{L=1}^{2}{\hat{n}_{Ld\uparrow} \hat{n}_{Ld\downarrow}} \\
    & + U_{pd}\sum_{L=1}^{2}{\hat{n}_{Ld}\hat{n}_p} + U_p\hat{n}_{p\uparrow}\hat{n}_{p\downarrow} + U_{dd}\hat{n}_{d1}\hat{n}_{d2}.
\end{aligned}
\label{H0}
\end{equation}
Here $\hat{n}_x$ denotes the electron number operator of spatial orbital $x$, 
and $n_{Ld}$ and $n_{Ld\uparrow}$ denote the spin-summed and spin-up electron number operator on $d$ orbital of Cu atom $L$, respectively. 
The perturbative part $\hat{H}_1$ is formulated as
\begin{equation}
\begin{aligned}
    \hat{H}_1 & = \left(\hat{h}_{d1,p}(-t_{pd1}(\hat{n}_{d1}-\frac{1}{2}) + t_{pd2}(\hat{n}_p-\frac{1}{2}) + t_{pd3}\hat{n}_{d2}) \right. \\
    & \left. - \hat{h}_{d2,p}(-t_{pd1}(\hat{n}_{d2}-\frac{1}{2}) + t_{pd2}(\hat{n}_p-\frac{1}{2}) + t_{pd3}\hat{n}_{d1})\right) \\
    & + \left(h.c.\right) + t_{pf} (\hat{h}_{f1,p} + \hat{h}_{f2,p}) + t_{dd} \hat{h}_{d1,d2} \\
    & + \frac{1}{2}K_{pd}\sum_{L=1}^2{\left(\hat{h}_{dL,p}^2-\hat{n}_{dL}-\hat{n}_{p}\right)} \\
    & + \frac{1}{2}K_{df}\sum_{L=1}^2{\left(\hat{h}_{dL,fL}^2-\hat{n}_{dL}-\hat{n}_{fL}\right)}, 
\end{aligned}
\label{H1}
\end{equation}
where $\hat{h}_{xy} = \sum_\sigma{(\hat{x}^\dagger_\sigma \hat{y_\sigma} + \hat{y}^\dagger_\sigma \hat{x_\sigma})}$ denotes the hopping between two spatial orbitals $x$ and $y$. 
$t_{pd1}$, $t_{pd2}$ and $t_{pd3}$ denote $dp$-hopping related to different integrals. 
The values of all parameters can be estimated from the \textit{ab initio} Hamiltonian with only the five orbitals corresponding to the model levels being active. 
The full table of parameters is presented in SI Table SVI. 
%
%
It turns out that $df$ angular exchange channel, represented by $K_{df}$, plays the dominant role for the Cu 4f contribution to $J$. 
For Cu 5p orbitals contribution, however, the $pf$ direct hopping surpasses the $df$ channel and becomes the leading force. 
%
In the following text, the contributions of $K_{df}$ and $t_{pf}$ to AFM $J$ will be considered separately. 
The rest of the Hamiltonian is a variant of 3-band model, which contains 
direct $dd$ hopping $t_{dd}$ and $dp$-exchange $K_{pd}$. 
We perform analysis of the 2-site $p$-$d$-$f$ model using the downfolding and perturbation method, which are explained in detail in SI Section \ref{Sec2_Downfold}.

Setting $|d_{1\uparrow}d_{2\downarrow}p^2\rangle$ and $|d_{1\downarrow}d_{2\uparrow}p^2\rangle$ to be the reference states, 
one can obtain the effective Hamiltonian on these states, and then the AFM coupling as the spin gap within the effective Hamiltonian. 
First, we constrain ourselves to the lowest order contribution to AFM coupling, the second-order with respect to $(H_1)$ ($(H_1)^2$-order) perturbation; magnetic coupling obtained this way is denoted as $J^{(2)}$.
It resembles the known result 
$J^{(2)} = {4t_{dd}^2}/{U_\text{eff}}$, where $U_\text{eff}$ is calculated as $U_\text{eff} = U_d-U_{dd} = E[d_1^2p^2] - E[d_{1\uparrow}d_{2\downarrow}p^2]$, and does not contain any $f$-level contribution to AFM coupling.
%
%
%

Moving to higher-order perturbations one by one, we find that 
%
$K_{df}$ contribution shows up only at the $(H_1)^4(\epsilon_f-\epsilon_d)^{-1}$ order.
If we adopt $J^{(m,n)}$ notation for $J$ obtained using the $(H_1)^m(\epsilon_f-\epsilon_d)^{-n}$-order perturbation, the leading contribution of $K_{df}$ to $J$ is $J^{(4,1)}_{K}$:
\begin{equation}
    J^{(4,1)}_{K} = \frac{2K_{df}^2t_{dd}^2}{(\epsilon_f-\epsilon_d)U_\text{eff}^2}.
\end{equation}
%
%
$J^{(4,1)}_{K}$ is always positive, which indicates the AFM contribution of Cu 4f to magnetic coupling.

The lowest-order occurrence of $t_{pf}$ is at $(H_1)^4(\epsilon_f-\epsilon_d)^{-2}$ and $(H_1)^5(\epsilon_f-\epsilon_d)^{-1}$ order. 
The former reads
\begin{equation}
    J^{(4,2)}_{t} = \frac{16t_{pf}^2t_{dd}^2}{(\epsilon_f-\epsilon_d)^2U_\text{eff}},
\end{equation}
while the latter is
\begin{equation}
    J^{(5,1)}_{t} = \frac{16t_{pf}^2t_{2}}{(\epsilon_f-\epsilon_d)U_\text{1CT}^3}\left(\frac{t_{dd}t_{1}U_\text{1CT}}{U_\text{eff}} + t_{pd2}(K_{pd}-t_{dd})\right),
\end{equation}
where $U_\text{1CT} = \epsilon_d-\epsilon_p+U_d-U_{pd}+U_{dd}+U_p = E[d_1^2d_{2\downarrow}p_{\uparrow}] -E[d_{1\uparrow}d_{2\downarrow}p^2]$, $t_{1} = \langle d_{1\downarrow}d_2^2p_\uparrow|\hat{H}|d_2^2p^2\rangle = 2t_{pd3} + t_{pd2}$, and $t_{2} = \langle d_{1\downarrow}d_{2\uparrow}p^2|\hat{H} |d_{1\downarrow}d_2^2p_\uparrow\rangle = t_{pd1} - t_{pd2} - t_{pd3}$.
Our numerical results show $K_{pd} > t_{dd}$, hence both $J^{(4,2)}_{t}$ and $J^{(5,1)}_{t}$ are positive contributions to $J$. 
This demonstrates the AFM contribution to $J$ arising from the Cu 5p orbitals.

In summary, based on the extended model, we find the increase of AFM $J$ due to the introduction of Cu 4f and 5p have different sources. 
Cu 4f orbitals are involved with $df$ angular exchange channel of superexchange characterized with $K_{df}$, which reduces the effective 3d on-site Hubbard repulsion by accepting electron-pair hopping from 3d to 4f. 
Among the Cu 4f shell, the $x(x^2-y^2)$, $y(x^2-y^2)$ and $z(x^2-y^2)$ components have the largest $K_{df}$ integrals ($\sim 0.1\mathrm{~E_h}$) and contribute the most to the $df$ angular exchange channel. 
Cu 5p orbitals, however, do not have such large $K_{df}$ integrals, and participate in AFM coupling via $pf$ direct hopping characterized with $t_{pf}$. 
%
%
%
%
The presented $p$-$d$-$f$ model can be further employed in future theoretical research to encode electron correlations within high-energy bands, improving the low-energy models and our understanding of high-temperature superconductivity.

\section*{Discussion}\label{Discussion}

By leveraging accurate correlated WFT calculations, we have comprensensively elucidated the orbtial-resolved contributions to the superexchange mechanism in cuprates. 
The exceptionally strong AFM coupling prevalent in these materials has been demonstrated to predominantly originate from electronic correlations within the high-energy copper bands. 
This correlation-driven enhancement comprises three synergistic components: 
(i) The \textit{radial breathing} effect, manifested in high-energy Cu d orbitals such as 4d and 5d, arises from radial hybridization between Cu 3d orbitals. 
(ii) The \textit{angular exchange} effect, prominent in Cu 4f orbitals, stems from exchange interactions between d-orbitals and those with distinct angular momentum symmetries;
(iii) The \textit{direct hopping} effect, observed in Cu 5p orbitals, emerges from oxygen-mediated hopping processes involving high-energy Cu orbitals. 
Collectively, these mechanisms magnify the Cu-centered AFM exchange interaction from an unrenormalized value of $\mathrm{\sim 35~meV}$ to $\mathrm{\sim 127~meV}$, establishing the microscopic foundation for the extraordinary spin-fluctuation-mediated high-temperature superconductivity. 

Our quantitative analysis further elucidates the collective effects of high-energy atomic shells on magnetic coupling.
The collective effects can be categorized into individual virtual contributions (IVC, 31 meV), core-virtual cooperative contributions (CV-CC, 18 meV), and environmental contributions (2 meV), according to the orbitals excitations involved.
%
The aforementioned mechanisms driven by copper orbitals primarily operate through individual orbital channels, while cooperative effects are manifested primarily in O 3d orbitals and highly delocalized Cu/La states. 
These computational insights reveal the atomic-scale origins of dominant electronic correlation effects on magnetic coupling, offering a refined perspective on the physics of cuprates. 

Significantly, this work uncovers a more comprehensive picture of high-energy orbital correlation effects on enhancing the nearest neighbor AFM coupling. 
This achievement not only provides guidance to design appropriate active spaces for subsequent theoretical calculations, but also identifies the physically most crucial channels in superexchange formation. 
Our study accentuates the significance of high-energy Cu orbitals within the low-energy physics of cuprates, paving the way for improving commonly used theoretical models.
The rectification of high-energy copper bands in low-energy models may encompass the following facets: (i) amendments to effective parameters in existing 3-band models; (ii) expansions of the model to incorporate $p$-$d$ bands; (iii) extensions of the model to account for high-energy bands.
All these aspects can potentially tailor the model phase diagram behavior in a quantitative or qualitative way. 
This study also lay a solid foundation to further explore the dynamic correlation effects beyond AFM coupling in cuprates.
Resolving the correlation effects on doped states or longer-range magnetic coupling will further advance our understanding of relationship between chemical composition and physical properties of high-temperature superconductors. 
%

%
%
%
\section*{Methods} \label{Methods}

The crystal structure of $\mathrm{La_2CuO_4}$ is taken from Crystallography Open Database No. 2002183 \cite{Grazulis_COD_2012,Grande_La2CuO4Structure_1977}, with an orthorhombic $Abma$ symmetry and crystal constants $a=\mathrm{5.406\AA}, b=\mathrm{5.370\AA}, c=\mathrm{13.15\AA}$ (Fig. \ref{Fig1_structure}a).
In our embedding computation scheme, the $\mathrm{La_2CuO_4}$ crystal is divided into 3 layers (Fig. \ref{Fig1_structure}e): core layer (A) ($\mathrm{Cu_2O_{11}La_4}$) handled with high-order wave function theories,  middle layer (B) ($\mathrm{Cu_6O_{16}La_{12}}$) handled with spin-averaged Hartree-Fock, and outermost layer (C) handled as an array of point charges. 

\begin{figure*}
    \centering
    \includegraphics[width=16.0cm]{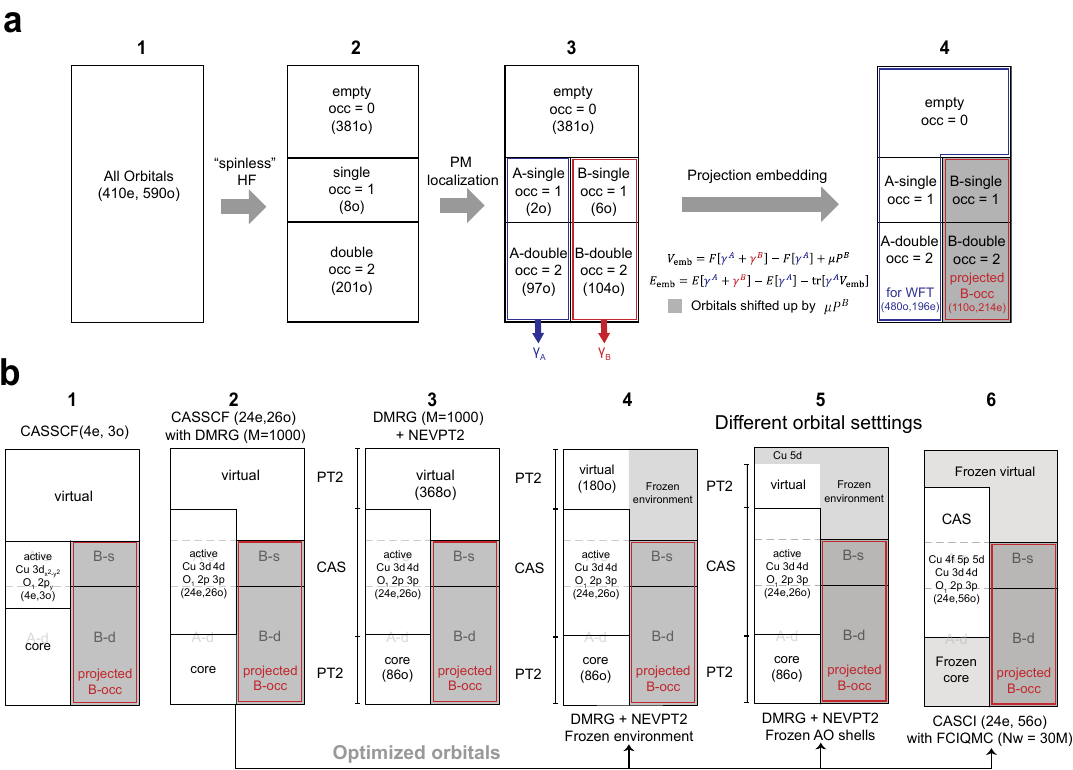}
    \caption{
    Computational workflow and settings. 
    (a) Quantum embedding scheme. 
    A 3-layer spin-averaged Hartree-Fock calculation partitions the entire orbital space into 3 parts: empty, singly occupied and doubly occupied orbitals. 
    Pipek-Mezey (PM) localization method is then used to partition singly and doubly occupied orbitals into regions A (blue rectangle) and B (red rectangle), from which RDMs of A and B ($\gamma_A$ and $\gamma_B$) are obtained. 
    $\gamma_A$ and $\gamma_B$ are then used to construct embedding potential $V_\text{emb}$ and $E_\text{emb}$.
    B occupied orbitals are projected out of WFT calculation by shifting their energy by a large positive value $\mu$, while the remaining orbital space (blue polygon) is considered for subsequent WFT calculations. 
    (b) Correlated WFT calculations using
    CASSCF(4e,3o), DMRG+CASSCF(24e,26o), NEVPT2 and FCIQMC. 
    Dark grey color with red boundary shows the projected occupied orbitals of region B. 
    The ``core'', ``active'' and ``virtual'' denote the doubly occupied, correlated and empty orbitals, respectively. 
    In NEVPT2 calculations, these terms mean perturbative core orbitals, multi-reference wave function space and perturbative virtual orbitals, respectively. 
    Light grey shades indicate the frozen virtual/core orbitals in NEVPT2 and FCIQMC calculations. 
    }
    \label{Fig4_method}
\end{figure*}

\subsection*{Point charge embedding}

As the first step of our embedding procedure, an array of point charges located outside of A+B (quantum cluster) is generated to reproduce the electrostatic potential within the quantum cluster with the \texttt{chargedel} tool, using the extended Evjen scheme \cite{gelle_evjen_2008,bogdanov_chargedel}.
The $\mathrm{Cu_{8}O_{27}La_{16}}$  quantum cluster is centered at two nearest neighbor copper sites and contains all atoms adjacent to these sites (Fig. \ref{Fig1_structure}e). 
The electrostatic potential within the quantum region is approximated to reproduce the Madelung potential, assuming that La, Cu and O have formal charges of $+3$, $+2$, $-2$ valence. 
The point charges are further divided into 2 parts, the normal charges and the scaled charges. 
The normal charges are point charges placed at the atom sites close to the quantum cluster with the formal charge values. 
The scaled charges are placed at the atom sites farther away from quantum cluster with charges scaled according to \cite{gelle_evjen_2008} to ensure fast convergence to the Madelung potential. 
%


%
%
%
%

%
%

\subsection*{Projection embedding calculation}

A smaller cluster, $\mathrm{Cu_{2}O_{11}La_{4}}$, is selected out of the center of quantum cluster as WFT cluster (A) to host the correlation treatment (Fig. \ref{Fig1_structure}e). 
The cluster A consists of a Cu-O-Cu structure, along with its 6 peripheral oxygen, 4 apical oxygen and 4 shoulder lanthanum atoms (Fig. \ref{Fig1_structure}d). 
All Cu and O atoms within A are treated with cc-pVDZ basis set (Cu: 6s5p3d1f; O: 3s2p1d) \cite{dunning_ccpvdz_1989a,balabanov_ccpvdz_2006a}, while La atoms are treated with a 2s2p1d basis set selected out of ECP46MWB with ECP \cite{Dolg_ECPMWB_1993}. 
B layer atoms are treated with smaller basis sets where more core electrons are represented with ECP \cite{hurley_CrenbsCu_1986a,ross_CrenbsLa_1990a}, whose details can be found in SI. 
%
%
The entire basis set of A+B has 590 basis functions and 410 electrons (Fig. \ref{Fig4_method}a step 1). 

The projection embedding scheme 
is followed to obtain an embedded Hamiltonian of A to describe its local electronic structure (Fig. \ref{Fig4_method}a), similar to procedure in Ref \cite{Manby_ProjEmb_2012}.
The basic idea of projection embedding, in the case of WFT in HF, is to freeze the buffer wave function to the HF result. 
One starts from a converged HF solution for the whole quantum cluster (Fig. \ref{Fig4_method}a step 2), partitions it into a direct product of cluster (A) and buffer (B), and then change A wave function to a correlated one, but leave B wave function unchanged. 
The freezing of B wave function is achieved in two steps: (1) by adding the Coulomb and exchange interaction between A and B into the external potential experienced by A, and (2) by increasing the energies of B-occupied orbitals by a large value, to effectively exclude those from subsequent calculations.
%

%
In cuprates, however, the ground state of parent compounds, even of a local cluster, is an antiferromagnetic state where spin-1/2 sites of A are entangled with those of B, such that A+B cannot be approximated by a product spin-adapted HF wave function. %
%
Therefore, in this work, a modified three-layer HF scheme, termed ``spin-averaged HF'', is used to obtain a spin-averaged open-shell environment for the cluster A, which preserves the direct product separability. 
In the following sub-sections, we will introduce the original projection embedding method, its spin-averaged HF variant, and the localization method in the presence of the spin-averaged HF.

\subsubsection*{Original projection embedding method}

Let the quantum cluster HF wave function be $|\Psi^0\rangle$, and assume that it can be partitioned into a direct product of A and B. 
\begin{equation}
    |\Psi^0\rangle = \hat{A}_0^\dagger\hat{B}_0^\dagger|\mathrm{vac}\rangle
\end{equation}
The correlated methods are used to get a $\hat{A}^\dagger$ to describe A better. 
The embedded Hamiltonian is defined as 
\begin{equation}
    \langle\Psi_A^1|\hat{H}_\text{emb}|\Psi_A^2\rangle = \langle\Psi_A^1|\hat{B}_0\hat{H}\hat{B}_0^\dagger|\Psi_A^2\rangle
\end{equation}
In this embedded Hamiltonian, the density matrix of B subsystem serves as a parameter, which is formulated as 
\begin{equation}
    (\rho^0_B)_{pq} = \langle\text{vac}|\hat{B}_0\hat{p}^\dagger\hat{q}\hat{B}_0^\dagger|\text{vac}\rangle
\end{equation}
Then the embedding Hamiltonian can be formulated as
\begin{equation}
\begin{aligned}
    & \langle\Psi_A^1|\hat{H}_\text{emb}|\Psi_A^2\rangle & \\
    & = \langle\Psi_A^1|\hat{B}_0 \big[\hat{H}_A + \hat{H}_B \\
    & + \sum_{a_1a_2b_1b_2}{\big((a_1a_2|b_1b_2) - (a_1b_2|b_1a_2)\big)\hat{a}_1^\dagger\hat{b}_1^\dagger\hat{b}_2\hat{a}_2} \big] \hat{B}_0^\dagger|\Psi_A^2\rangle \\
    & = \langle\Psi_A^1|\big[\hat{H}_A + E^0_B \\
    & + \sum_{a_1a_2b_1b_2}{\big((a_1a_2|b_1b_2) - (a_1b_2|b_1a_2)\big)\hat{a}_1^\dagger\hat{a}_2(\rho^0_B)_{b_1b_2}}\big]|\Psi_A^2\rangle \\
\end{aligned}
\end{equation}
The 1-body embedding potential is 
\begin{equation}
\begin{aligned}
    \hat{V}_\text{emb} & = \sum_{a_1a_2b_1b_2}{\big((a_1a_2|b_1b_2) - (a_1b_2|b_1a_2)\big)\hat{a}_1^\dagger\hat{a}_2}(\rho_B^0)_{b_1b_2} \\
    & = (F[\rho_A^0 + \rho_B^0] - F[\rho_B^0])_{a_1a_2}\hat{a}_1^\dagger\hat{a}_2.  \\   
\end{aligned}
\label{Vemb_potential}
\end{equation}
The energy shift is
\begin{equation}
\begin{aligned}
    E_\text{emb} & = E_B^0 = E[\rho_A^0 + \rho_B^0] - E[\rho_A^0] \\
    & - \sum_{a_1a_2}{(F[\rho_A^0 + \rho_B^0] - F[\rho_B^0])_{a_1a_2}(\rho^0_A)_{a_1a_2}}. 
\end{aligned}
\label{Eemb_potential}
\end{equation}
$F[\cdot]$ and $E[\cdot]$ denote Fock matrix and energy functional of Hartree-Fock type, respectively. 
These additional terms are called embedding potential and energy correction, respectively. 
However, the embedding potential cannot restrict A wave function onto A orbitals, since B's occupied orbitals are still of low energy. 
Therefore, an energy shifting term (``projection term'') is added to the embedding potential to raise the B occupied orbital energy to a prohibitively high level, forbidding A electrons to interact with them (Fig. \ref{Fig4_method}a step 3 and 4).
\begin{equation}
    \hat{P}_B = \mu\sum_b{\hat{b}^\dagger\hat{b}}
\end{equation}
In this work, we set $\mu = \mathrm{10^5 E_h}$. 

In this way, we obtain the embedding potential $\hat{V}_\text{emb} + \hat{P}_B$ and the energy correction $\hat{E}_\text{emb}$, which describes the effect of environment B on A with B's density matrix as the only requirement. 

\subsubsection*{Spin-averaged HF variant of projection embedding}

In cuprates, the AFM ground state cannot be expressed with conventional RHF or ROHF wave functions, even approximately. 
In order to describe AFM environment at the mean-field level, we proposed a revised version of Hartree-Fock theory named spin-averaged HF to settle this problem.
This procedure is similar to configuration-averaged Hartree–Fock (CAHF) \cite{mcweeny_cahf_1974,zerner_CAHF_1989}.

The spin-1/2 sites form AFM ground state, which can neither be separated into a direct product of A and B, nor expressed as a Slater determinant (SD), even in an approximate way. 
A direct localization of RHF orbitals requires ad-mixture between occupied and unoccupied orbitals, which changes the RDM. 
However, following smearing methods of SCF, we can define a Fock matrix with spinless fractional occupation. 
\begin{equation}
    F_{pq} = h_{pq} + \sum_r{f_r\left((pq|rr) - \frac{1}{2}(pr|rq)\right)}, 
    f_r = \left\{
    \begin{array}{lr}
        2, & r \in d; \\
        1, & r \in s; \\
        0, & r \in u. 
    \end{array}\right.
\end{equation}
where $p$, $q$ and $r$ denote spatial MOs, and $f_r$ denote the occupation number of $r$, and $d$, $s$, $u$ denote doubly-occupied, singly-occupied and empty orbitals, respectively (Fig. \ref{Fig4_method}a step 2).
In our system, the orbital occupation numbers are set according to the orbital energy order, and the number of Cu $\mathrm{3d}_{x^2-y^2}$ open-shell orbitals of the system (Fig. \ref{Fig1_structure}d). 
SCF calculations can be done to diagonalize this Fock matrix, which yields a mean field approximation to the AFM ground state of the system, as well as a set of well-behaved orbitals as a starting point for further analysis. 
Since this HF method describes open-shell systems by considering the average effect of different spin configurations of environment, rather than a specific spin configuration, we dub this scheme ``spin-averaged HF''.

\subsubsection*{Localization method}

$d_A$, $d_B$, $s_A$, $s_B$ denote the intersection between doubly (singly) occupied orbitals and A (B) subsystem, and subsystem orbital set is defined by PM localization method (Fig. \ref{Fig4_method}a step 3). 
This density matrix is then substituted into Eq (\ref{Vemb_potential}) and Eq (\ref{Eemb_potential}) to get the 1-body and 0-body embedding potential.

After delocalized spin-averaged HF orbitals are constructed, the (spin-traced) density matrix by mixed-HF is 
\begin{equation}
    (\rho^0)_{pq} = f_p \delta_{pq},
\end{equation}
Pipek-Mezey (PM) localization methods are used in doubly- ($d$) and singly-occupied ($s$) subspaces, respectively, to get localized orbitals without mixing orbitals of different occupation number. 
The subsystem partition of each local orbital is determined by the atom closest to it. 
If the closest atom to an orbital is a $\mathrm{Cu_2O_{11}La_4}$ (A) atom, then the orbital is labelled as A. 
If a $\mathrm{Cu_6O_{16}La_{12}}$ (B) atom is the closest, then the orbital is labelled as B. 
Then the total spin-averaged HF density matrix is partitioned into subsystem A and B, defined as follows. 
\begin{equation}
\begin{aligned}
    (\rho_A^0)_{pq} & = \left\{
    \begin{array}{lr}
        2\delta_{pq}, & p \in d \cap \text{A}; \\
        \delta_{pq}, & p \in s \cap \text{A}; \\
        0, & p \in u \cup \text{B}. 
    \end{array}\right. \\
    (\rho_B^0)_{pq} & = \left\{
    \begin{array}{lr}
        2\delta_{pq}, & p \in d \cap \text{B}; \\
        \delta_{pq}, & p \in s \cap \text{B}; \\
        0, & p \in u \cup \text{A}. 
    \end{array}\right. \\
\end{aligned}
\end{equation}

$\rho_A^0$ and $\rho_B^0$ is then substituted into Eq (\ref{Vemb_potential}) and Eq (\ref{Eemb_potential}) for the embedding potential (Fig. \ref{Fig4_method}a step 4). 

\subsection*{Correlated WFT calculations}

After the embedding potential is obtained, different WFT calculations are conducted to calculate the correlation effects on the magnetic coupling $J$ at different levels of approximation. 
The WFT calculation methods are illustrated in Fig. \ref{Fig4_method}b, and described as follows. 

\subsubsection*{CASSCF with DMRG as FCI solver (DMRG-CASSCF)}

As the first step, CASSCF calculation on a small CAS (4e,3o) is performed with exact diagonalization FCI solver with \texttt{PySCF}\cite{sun_libcint_2015,Sun_PySCF_2020} (Fig. \ref{Fig4_method}b step 1). 
The CAS consists of Cu $\mathrm{3d}_{x^2-y^2}$ and bridging O $\mathrm{2p}_\sigma$, and atomic valence active space (AVAS) \cite{Sayfutyarova_AVAS_2017} technique is used to generate the initial guess for orbital optimization. 
Then CASSCF on CAS(24e,26o) is performed with DMRG as the FCI solver.
DMRG is performed using the \texttt{BLOCK2} package \cite{Zhai_Block2_DMRG_2023}, with \texttt{PySCF} as the CASSCF driver\cite{Sun_PySCF_2020} (Fig. \ref{Fig4_method}b step 2). 
The DMRG calculation is performed with bond dimension $M=1000$. 
The enlarged CAS includes the full Cu 3d space and bridging O 2p space, as well as higher-energy Cu 4d and bridging O 3p. 
In both CASSCF(4e,3o) and CASSCF(24e,26o), the singlet and triplet states are converged. 
The converged orbital set from CASSCF(24e,26o) is used in the subsequent steps. 

\subsubsection*{Multi-reference perturbation (MRPT)}

After DMRG-based orbital optimization, the effects of higher-energy orbitals are studied by correlating high-energy orbitals with the strongly contracted second-order \textit{n}-electron valence state perturbation theory (SC-NEVPT2) with compressed perturber on the basis of DMRG (Fig. \ref{Fig4_method}b step 3).  \cite{Angeli_NEVPT2_2002,Guo_DMRG_SC_NEVPT_2016,Sokolov_DMRG_TD_NEVPT2_2017}
The NEVPT2 calculations can be split into three parts. 
First, the whole cluster space (blue polygon in Fig. \ref{Fig4_method}a step 4) is included in PT space to calculate the dynamic correlation effect of all high-energy orbitals. 
Next, the environment empty orbitals are removed from PT space to calculate the environment effects (Fig. \ref{Fig4_method}b step 4). 
Finally, multiple PT space settings are used to figure out the individual effects of each orbital shells (Fig. \ref{Fig4_method}b step 5). 
Bond dimension $M=1000$ and CAS(24e,26o) are used in all CAS and PT space settings. 
The atomic orbital shells in step 4 and 5 are defined by orthogonal projection of meta-L\"{o}wdin atomic orbitals \cite{Sun_metaLowdin_2014} onto the core or virtual space, followed by a Gram-Schmidt orthogonalization to preserve the orbitals' similarity to meta-L\"{o}wdin AOs.

\subsubsection*{Full configuration interaction quantum Monte Carlo (FCIQMC)}

In order to examine the accuracy of NEVPT2, FCIQMC calculations are performed and compared with NEVPT2 results on several small PT spaces (Fig. \ref{Fig4_method}b step 6) \cite{booth_origFCIQMC_2009,guther_neci_2020}. 
To improve convergence and reduce computational errors, we use the initiator approximation and adaptive shift methods throughout this work, with an initiator threshold $n_a = 3$ \cite{cleland_iFCIQMC_2010,ghanem_aaFCIQMC_2019}. 
We use a semi-stochastic approach with a deterministic space size of 1000 \cite{blunt_semiStochastic_2015}.
To obtain statistical average of energy, trial wave functions are constructed by diagonalizing the Hamiltonian within a subspace spanned by 10 most occupied SDs. 
In order to eliminate the initiator errors, we test FCIQMC with varying numbers of walkers, reaching up to 100 million walkers, such that the systematic error in energy is reduced to below 2 meV, which is very close to the typical statistic error in these systems. 
FCIQMC calculations on several typical active spaces involved in AFM $J$ are performed to check their perturbation error. 
The FCIQMC results are presented in SI Section IV.

\subsubsection*{WFT evaluation of AFM coupling $J$}

To compute AFM coupling $J$, we can follow the standard routine, where the energy spectrum of the embedded cluster (
core layer A), containing 2 copper atoms, is mapped to a 2-site $S=1/2$ nearest-neighbor Heisenberg model \cite{de_graaf_magnetic_2016}.
\begin{equation}
    \hat{H} = J\hat{\mathbf{S}}_1\cdot\hat{\mathbf{S}}_2.
\end{equation}
In the AFM coupling case ($J>0$), the ground state and the first excited state of the model above are the singlet and triplet states: 
\begin{equation}
\begin{aligned}
    |\Psi_0\rangle=\frac{1}{\sqrt{2}}(|\uparrow\downarrow\rangle - |\downarrow\uparrow\rangle), & \quad E_0 = -\frac{3J}{4}; \\
    |\Psi_1\rangle=\frac{1}{\sqrt{2}}(|\uparrow\downarrow\rangle + |\downarrow\uparrow\rangle), & \quad E_1 = \frac{J}{4}. 
\end{aligned}
\end{equation}
Therefore, the lowest-energy spin singlet and triplet states in \textit{ab initio} calculation are mapped to $|\Psi_0\rangle$ and $|\Psi_1\rangle$ above, and $J$ is calculated as the difference between corresponding energies. 
The spin-1/2 on each site comes mainly from electrons occupying the open-shell Cu $\mathrm{3d}_{x^2-y^2}$ orbitals.
However, the many-body wave functions of low-energy states also include electron correlations from other high-energy orbitals, which significantly contribute to the magnetic coupling.

\begin{acknowledgments}
This work was supported by the National Key R\&D Program of China under Grant No. 2021YFA1400500, 
the Strategic Priority Research Program of the Chinese Academy of Sciences under Grant No. XDB33000000, 
the National Natural Science Foundation of China under Grant No. 12334003, 
and the Beijing Municipal Natural Science Foundation under Grant No. JQ22001. 
We are grateful for computational resources provided by the High Performance Computing Platform of Peking University. 
\end{acknowledgments}

\bibliographystyle{apsrev4-1}
\def\bibfont{\footnotesize}
\bibliography{main}

\clearpage
\onecolumngrid

\newcommand{\mainmatter}{}
%
%
\newif\ifmain
\ifx\mainmatter\undefined
  \mainfalse
\else
  \maintrue
\fi

\ifmain

\setcounter{page}{0}
\setcounter{figure}{0}
\setcounter{equation}{0}
\setcounter{table}{0}

\renewcommand{\thepage}{S\arabic{page}}
\makeatletter 
\renewcommand{\thefigure}{S\@arabic\c@figure}
\makeatother

\makeatletter 
\renewcommand{\theequation}{S\@arabic\c@equation}
\makeatother

\makeatletter 
\renewcommand{\thetable}{S\@Roman\c@table}
\makeatother

\thispagestyle{empty}
{\centering
\LARGE \textbf{Supplementary Information: Individual and cooperative superexchange enhancement in cuprates} \\[1em]
\normalsize
Tonghuan Jiang\textsuperscript{1}, Nikolay A. Bogdanov\textsuperscript{2}$^*$, Ali Alavi\textsuperscript{2,3}$^\dagger$, Ji Chen\textsuperscript{1,4,5}$^\ddagger$ \\[0.5em]
\footnotesize
\textsuperscript{1}School of Physics, Peking University, Beijing 100871, P. R. China \\
\textsuperscript{2}Max Planck Institute for Solid State Research, Heisenbergstrasse 1, 70569 Stuttgart, Germany \\
\textsuperscript{3}University of Cambridge, Lensfield Road, Cambridge CB2 1EW, United Kingdom \\
\textsuperscript{4}Interdisciplinary Institute of Light-Element Quantum Materials and Research Center for Light-Element Advanced Materials, Peking University, Beijing 100871, P. R. China \\
\textsuperscript{5}Frontiers Science Center for Nano-Optoelectronics, Peking University, Beijing 100871, P. R. China \\

\vspace{1em}
$^*$ n.bogdanov@fkf.mpg.de \\
$^\dagger$ a.alavi@fkf.mpg.de \\
$^\ddagger$ ji.chen@pku.edu.cn \\
}

\vspace{2em}
\section*{Contents}
I. Details of NEVPT2 calculations  \hfill \pageref{SecI_ShellChoice}

\vspace{1em}
II. L\"{o}wdin downfolding for perturbation analysis  \hfill \pageref{Sec2_Downfold}

\vspace{1em}
III. Perturbation theory analysis on roles of Cu 4f and 5d  \hfill \pageref{Sec3_PerturbAnalysis}

\vspace{1em}
IV. FCIQMC results  \hfill \pageref{SecIV_FCIQMC}

\vspace{1em}
V. Cluster structure and basis sets  \hfill \pageref{SecV_structure}

\vspace{1em}

\clearpage

\else

\documentclass[reprint,superscriptaddress,amsmath,amssymb,aps,prb,floatfix,onecolumn,a4paper]{revtex4-2}

\usepackage{multirow}
\usepackage{mlmodern}
\usepackage[utf8]{inputenc}
\usepackage{graphicx}
\usepackage{enumitem}
\usepackage{gensymb}
\usepackage{amsmath}
\usepackage{amsthm}
\usepackage{amssymb}
\usepackage{dcolumn}
\usepackage{color}
\usepackage{bbm}
\usepackage{verbatim}
\usepackage{hyperref}
\usepackage{longtable}
\usepackage[normalem]{ulem}
\usepackage{xr}
\makeatletter

\newcommand{\bc}[1]{{\color{blue} #1 }}

\newtheorem{theorem}{Statement}

\bibliographystyle{apsrev4-1}

\newcommand*{\addFileDependency}[1]{
\typeout{(#1)}
%
%
\@addtofilelist{#1}
%
\IfFileExists{#1}{}{\typeout{No file #1.}}
}\makeatother

\newcommand*{\myexternaldocument}[1]{%
\externaldocument{#1}%
\addFileDependency{#1.tex}%
\addFileDependency{#1.aux}%
}

\myexternaldocument{../main}

\begin{document}

\title{Supporting Information: Individual and cooperative superexchange enhancement in cuprates}


\author{Tonghuan Jiang}
\affiliation{School of Physics, Peking University, Beijing 100871, P. R. China}

\author{Nikolay A. Bogdanov}
\email{n.bogdanov@fkf.mpg.de}
\affiliation{Max Planck Institute for Solid State Research, Heisenbergstrasse 1, 70569 Stuttgart, Germany}

\author{Ali Alavi}
\email{a.alavi@fkf.mpg.de}
\affiliation{Max Planck Institute for Solid State Research, Heisenbergstrasse 1, 70569 Stuttgart, Germany}
\affiliation{University of Cambridge, Lensfield Road, Cambridge CB2 1EW, United Kingdom}

\author{Ji Chen}
\email{ji.chen@pku.edu.cn}
\affiliation{School of Physics, Peking University, Beijing 100871, P. R. China}
\affiliation{Interdisciplinary Institute of Light-Element Quantum Materials and Research Center for Light-Element Advanced Materials,Peking University, Beijing 100871, P. R. China
}
\affiliation{Frontiers
Science Center for Nano-Optoelectronics, Peking University, Beijing 100871, P. R. China}

\date{\today}

\makeatletter 
\renewcommand{\thefigure}{S\@arabic\c@figure}
\makeatother

\makeatletter 
\renewcommand{\theequation}{S\@arabic\c@equation}
\makeatother

\makeatletter 
\renewcommand{\thetable}{S\@Roman\c@table}
\makeatother

\maketitle

\tableofcontents
\clearpage

\fi

\begin{table}[h!]
\normalsize
\section{Details of NEVPT2 calculations}\label{SecI_ShellChoice}
\raggedright
In this section, a list of all PT space used in this work is presented, along with $J$ values. It is also shown here from which NEVPT2 calculations the contributions of each orbital are obtained.  
\caption{
\label{Table1_longlong}
DMRG+NEVPT2 calculations, from which the contribution of environment, core, virtual and each atomic shell are extracted. 
``PT core space'' and ``PT virtual space'' denote the perturbative core and virtual space used in NEVPT2 calculation, whose sizes are marked in the ``PT size'' column. 
$(m,n)$ denotes $(m\text{c},n\text{v})$, i.e. $m$ core orbitals and $n$ virtual orbitals are correlated with NEVPT2. 
\mbox{Low-E = O 3sp + Cu 4sp 5sp + La 5d}, 
\mbox{Mid-E = Cu 5d 6sp + La 6sp}, 
\mbox{High-E = O 3d + Cu 4f 6p}.
}
\begin{ruledtabular}
{\renewcommand{\arraystretch}{1.0}
\begin{tabular}{ccccd}

\textrm{Setting id}&
\textrm{PT core space}&
\textrm{PT virtual space}&
\textrm{PT size}&
\multicolumn{1}{c}{\textrm{$J$ (meV)}}\\
\hline
a1 & None & None & (0,0) & 92.24 \\
a2 & Full core & Full virt + env & (86,368) & 143.78 \\
a3 & Full core & Full virt & (86,180) & 141.73 \\
a4 & Full core & None & (86,0) & 91.82 \\
a5 & None & Full virt & (0,180) & 123.74 \\
\hline
b1 & Cu 3, La 5, O 2 & Full virt & (65,180) & 141.39 \\
b2 & Cu 3, O 2 (Core-CuO) & Full virt & (49,180) & 141.79 \\
b3 & Core-CuO w/o Cu 3s & Full virt & (47,180) & 139.24 \\
b4 & Core-CuO w/o Cu 3p & Full virt & (43,180) & 139.25 \\
b5 & Core-CuO w/o O$_b$ 2s & Full virt & (48,180) & 144.33 \\
b6 & Core-CuO w/o O$_a$ 2 & Full virt & (33,180) & 139.07 \\
b7 & Core-CuO w/o O$_p$ 2 & Full virt & (25,180) & 130.30 \\
b8 & Core-CuO w/o O$_p$ 2s & Full virt & (43,180) & 136.47 \\
b9 & Core-CuO w/o O$_p$ 2p & Full virt & (31,180) & 130.66 \\
\hline
c1 & Core-CuO & LowE + MidE & (49,105) & 129.68 \\
c2 & Core-CuO & Full w/o O$_b$ 3d & (49,175) & 141.01 \\
c3 & Core-CuO & Full w/o O$_a$ 3d & (49,160) & 141.50 \\
c4 & Core-CuO & Full w/o O$_p$ 3d & (49,150) & 139.89 \\
c5 & Core-CuO & Full w/o Cu 6p & (49,174) & 140.25 \\
c6 & Core-CuO & Full w/o Cu 4f & (49,166) & 133.58 \\
c7 & Core-CuO & LowE + MidE + Cu 4f & (49,119) & 137.60 \\
c8 & Core-CuO & LowE + MidE + Cu 4f w/o Cu 6s & (49,117) & 136.06 \\
c9 & Core-CuO & LowE + MidE + Cu 4f w/o Cu 5d & (49,109) & 120.95 \\
c10 & Core-CuO & LowE + MidE + Cu 4f w/o La 6sp & (49,103) & 136.21 \\
c11 & Core-CuO & LowE + Cu 4f 5d & (49,101) & 134.53\\
c12 & Core-CuO & LowE + Cu 4f 5d w/o La 5d & (49,81) & 130.54 \\
c13 & Core-CuO & LowE + Cu 4f 5d w/o Cu 5s & (49,99) & 133.84 \\
c14 & Core-CuO & LowE + Cu 4f 5d w/o Cu 5p & (49,95) & 122.97 \\
c15 & Core-CuO & LowE + Cu 4f 5d w/o Cu 4s & (49,99) & 135.04 \\
c16 & Core-CuO & LowE + Cu 4f 5d w/o Cu 4p & (49,95) & 130.86 \\
c17 & Core-CuO & LowE + Cu 4f 5d w/o O$_b$ 3s & (49,100) & 137.13 \\
c18 & Core-CuO & LowE + Cu 4f 5d w/o O$_a$ 3sp & (49,85) & 134.07 \\
c19 & Core-CuO & LowE + Cu 4f 5d w/o O$_p$ 3sp & (49,77) & 130.74 \\
c20 & Core-CuO & Cu 4f 5d 5p & (49,30) & 127.65 \\
c21 & Core-CuO & None & (49,0) & 93.02 \\
c22 & None & Cu 4f 5d 5p & (0,30) & 127.70 \\
\hline
d1 & Core-CuO & Cu 4pf 5pd O$_{b,p}$ 3sp & (49,61) & 130.14 \\
d2 & Core-CuO & Cu 4pf 5pd O$_{b,p}$ 3sp La 5d & (49,81) & 134.20 \\
d3 & Core-CuO & Cu 4spf 5spd (45) O$_{b,p}$ 3sp La 5d & (49,85) & 134.07 \\
d4 & Core-CuO & Cu 45 O 3sp La 5d & (49,101) & 134.53 \\
d5 & Core-CuO & Cu 45 O 3sp La 56 & (49,117) & 136.06 \\
d6 & Core-CuO & Cu 456 O 3sp La 56 & (49,125) & 139.13 \\
d7 & Core-CuO & Cu 456 O 3sp O$_b$ 3d La 56 & (49,130) & 139.70 \\
\hline
e1 & None & Cu 4pf 5pd O$_{b,p}$ 3sp & (0,61) & 130.11 \\
e2 & None & Cu 4pf 5pd O$_{b,p}$ 3sp La 5d & (0,81) & 130.86 \\
e3 & None & Cu 4spf 5spd (45) O$_{b,p}$ 3sp La 5d & (0,85) & 131.21 \\
e4 & None & Cu 45 O 3sp La 5d & (0,101) & 131.38 \\
e5 & None & Cu 45 O 3sp La 56 & (0,117) & 131.68 \\
e6 & None & Cu 456 O 3sp La 56 & (0,125) & 131.83 \\
e7 & None & Cu 456 O 3sp O$_b$ 3d La 56 & (0,130) & 123.26 \\
\end{tabular}
}
\end{ruledtabular}
\end{table}
\begin{table}[h!]
\normalsize
\caption{\label{J_whole_SI}%
\normalsize
AFM coupling $J$ contribution of core, virtual and environment orbitals as a whole (major shell). 
The term \mbox{``A with (w/o) B''} means the contribution of A in the presence (absence) of B's correlation. 
($m$,$n$) in the ``PT size'' column means that $m$ core orbitals and $n$ virtual orbitals are involved in this major shell. 
The ``Source'' column shows the setting IDs (see ``Setting id'' column of Table \ref{Table1_longlong}) from which the contribution of this major shell are calculated. 
The cooperation between core and virtual is also shown in this table. 
%
%
}
\begin{ruledtabular}
{\renewcommand{\arraystretch}{1.05}
\begin{tabular}{cccd}
\textrm{Major Shell}&
\textrm{PT size}&
\textrm{Source}&
\multicolumn{1}{c}{\textrm{$\Delta J$ (meV)}} \\
\hline
Environment & (0,188) & a2 $-$ a3 & 3.05 \\
Core with Virtual & (86,0) & a3 $-$ a5 & 17.99 \\
Core w/o Virtual & (86,0) & a4 $-$ a1 & -0.42 \\
Virtual with Core & (0,180) & a3 $-$ a4 & 49.91 \\
Virtual w/o Core & (0,180) & a5 $-$ a1 & 31.50 \\
\hline
Cooperation between Core and Virtual & \textemdash & $(\text{a3}-\text{a5})-(\text{a4}-\text{a1})$ & 18.41 \\
\end{tabular}
}
\end{ruledtabular}
\end{table}
\begin{table}[h!]
\normalsize
\caption{\label{J_core_SI}%
\normalsize
The contribution of different core shells, along with the computations from which the contributions are extracted. 
$m$ in ``PT size'' column means that $m$ core orbitals are involved in this shell. 
``Source'' column shows the setting IDs (See ``Setting id'' column of Table \ref{Table1_longlong}) from which the contribution of this shell are calculated. 
%
%
}
\begin{ruledtabular}
{\renewcommand{\arraystretch}{1.05}
\begin{tabular}{cccd}
\textrm{Shell name}&
\textrm{PT size}&
\textrm{Source}&
\multicolumn{1}{c}{\textrm{$J$ (meV)}}\\
\colrule
Deep core & 21 & a3 $-$ b1 & 0.34 \\
La 5s 5p & 16 & b1 $-$ b2 & -0.40 \\
Cu 3s & 2 & b2 $-$ b3 & 2.55 \\
Cu 3p & 6 & b2 $-$ b4 & 2.54 \\
O$_1$ 2s & 1 & b2 $-$ b5 & -2.54 \\
O$_2$ 2s 2p & 16 & b2 $-$ b6 & 2.72 \\
O$_3$ 2s 2p & 24 & b2 $-$ b7 & 11.49 \\
O$_3$ 2s & 6 & b2 $-$ b8 & 5.32 \\
O$_3$ 2p & 18 & b2 $-$ b9 & 11.13 \\
\colrule
Total (Combined) & 86 & $\text{a3}-\text{a5}$ & 17.99 \\
Total (Summed) & 86 & $(\text{a3}-\text{b2})+(\text{b2}-\text{b3})+...+(\text{b2}-\text{b7})$ & 16.70 \\
\end{tabular}
}
\end{ruledtabular}
\end{table}

\begin{table}[h!]
\normalsize
\caption{\label{J_virt_SI}%
\normalsize
The contribution of different virtual shells, along with the computations from which the contributions are extracted. 
$n$ in ``PT size'' column means that $n$ virtual orbitals are involved in this shell. 
``Source'' column shows the setting IDs (See ``Setting id'' column of Table \ref{Table1_longlong}) from which the contribution of this shell are calculated. 
%
%
}
\begin{ruledtabular}
{\renewcommand{\arraystretch}{1.05}
\begin{tabular}{cp{25mm}<{\centering}p{40mm}<{\centering}d}
\textrm{Shell name} &
\textrm{PT size} &
\textrm{Source} &
\multicolumn{1}{c}{\textrm{$J$ (meV)}}\\
\colrule
O$_1$ 3d & 5 & $\text{b2}-\text{c2}$ & 0.78 \\
O$_2$ 3d & 20 & $\text{b2}-\text{c3}$ & 0.29 \\
O$_3$ 3d & 30 & $\text{b2}-\text{c4}$ & 1.90 \\
Cu 6p & 6 & $\text{b2}-\text{c5}$ & 1.54 \\
Cu 4f & 14 & $\text{b2}-\text{c6}$ & 8.21 \\
Cu 6s & 2 & $\text{c7}-\text{c8}$ & 1.54 \\
Cu 5d & 10 & $\text{c7}-\text{c9}$ & 16.65 \\
La 6s 6p & 16 & $\text{c7}-\text{c10}$ & 1.39 \\
La 5d & 20 & $\text{c11}-\text{c12}$ & 3.99 \\
Cu 5s & 2 & $\text{c11}-\text{c13}$ & 0.69 \\
Cu 5p & 6 & $\text{c11}-\text{c14}$ & 11.56 \\
Cu 4s & 2 & $\text{c11}-\text{c15}$ & -0.51 \\
Cu 4p & 6 & $\text{c11}-\text{c16}$ & 3.67 \\
O$_1$ 3s & 1 & $\text{c11}-\text{c17}$ & -2.60 \\
O$_2$ 3s 3p & 16 & $\text{c11}-\text{c18}$ & 0.46 \\
O$_3$ 3s 3p & 24 & $\text{c11}-\text{c19}$ & 3.79 \\
\colrule
Total (Combined) & 180 & $\text{b2}-\text{c21}$ & 48.77 \\
Total (Summed) & 180 & Sum of all rows above & 53.35 \\
\end{tabular}
}
\end{ruledtabular}
\end{table}
\clearpage

\begin{table}[h!]
\normalsize
\caption{\label{J_syner_SI}
\normalsize
The cooperative effects between core and virtual orbitals, which are evaluated by subtracting contribution of a virtual shell with core (column 2) with contribution without core (column 3). 
Information about the reference PT space are presented in the Supplementary Information Table \ref{Table1_longlong} 
%
%
}
\begin{ruledtabular}
{\renewcommand{\arraystretch}{1.2}
%
\begin{tabular}{cdcdcd}
\textrm{Virtual shell}&
\multicolumn{1}{c}{\textrm{Contribution with}} &
\textrm{Source with} &
\multicolumn{1}{c}{\textrm{Contribution}} &
\textrm{Source w/o} &
\multicolumn{1}{c}{\textrm{Synergistic effect}}\\
\textrm{name} $x$ &
\multicolumn{1}{c}{\textrm{core} $J[x;\text{core}]$} &
\textrm{ore} &
\multicolumn{1}{c}{\textrm{w/o core} $J[x]$} &
\textrm{core} &
\multicolumn{1}{c}{$J[x;\text{core}]-J[x]$}\\
\colrule
Cu 4f 5p 5d & 34.6 & $\text{c20}-\text{c21}$ & 35.5 & $\text{c22}-\text{a1}$ & -0.8 \\
O$_b$ 3s O$_p$ 3s 3p Cu 4p & 2.5 & $\text{d1}-\text{c20}$ & 2.4 & $\text{e1}-\text{c22}$ & 0.1 \\
La 5d & 4.1 & $\text{d2}-\text{d1}$ & 0.8 & $\text{e2}-\text{e1}$ & 3.3 \\
Cu 4s 5s & -0.1 & $\text{d3}-\text{d2}$ & 0.3 & $\text{e3}-\text{e2}$ & -0.5 \\
O$_a$ 3s 3p & 0.5 & $\text{d4}-\text{d3}$ & 0.2 & $\text{e4}-\text{e3}$ & 0.3 \\
La 6s 6p & 1.5 & $\text{d5}-\text{d4}$ & 0.3 & $\text{e5}-\text{e4}$ & 1.2 \\
Cu 6s 6p & 3.1 & $\text{d6}-\text{d5}$ & 0.2 & $\text{e6}-\text{e5}$ & 2.9 \\
O$_b$ 3d & 0.6 & $\text{d7}-\text{d6}$ & -8.6 & $\text{e7}-\text{e6}$ & 9.1 \\
O$_{(a,p)}$ 3d & 2.1 & $\text{b2}-\text{d7}$ & 0.5 & $\text{a5}-\text{e7}$ & 1.6 \\
\colrule
Total & 48.8 & $\text{b2}-\text{c21}$ & 31.5 & $\text{a5}-\text{a1}$ & 17.3 \\
\end{tabular}
}
\end{ruledtabular}
\end{table}

\section{L\"{o}wdin downfolding for perturbation analysis}\label{Sec2_Downfold}

Consider an interacting Hamiltonian in a large Hilbert space, where only a small partition of degrees of freedom (DOF) are of interest (denoted as 0), and the rest of DOF (denoted as 1) are supposed to have renormalization effect on the 0 subspace. 
In another words, the environment DOF (1) can be downfolded onto the DOF of interest (0). 
This includes two aspects: first, the correction on the effective Hamiltonian within the DOF of interest; second, the correction on the effective wave function of the DOF of interest. 

The total Hamiltonian is shown as follows, 
\begin{equation}
    H = \begin{pmatrix}
        H_{00} & H_{01} \\
        H_{10} & H_{11}
    \end{pmatrix},
\end{equation}
which consists of the unperturbed Hamiltonian $H_{00}$, the diagonal unperturbed Hamiltonian on the 1 space $H_{11}^{0}$, and the perturbation $H_{01}$ and $V_{11}$, with $H_{11}=H_{11}^{0}+V_{11}$. 
The effective Hamiltonian is formulated as 
\begin{equation}
    H_\mathrm{eff}(\omega) = H_{00} + H_{01}(\omega-H_{11})^{-1}H_{10}. 
    \label{Heff}
\end{equation}
where $\omega$ is the reference energy of downfolding, and is often taken as the energy of interest. 
As $\omega$ equals an eigenvalue of the full Hamiltonian, the eigenstates of effective Hamiltonian strictly yields the eigenstates of the total system projected onto 0 subspace. 
\begin{equation}
\begin{aligned}
    & \hat{H}|\Psi\rangle=E|\Psi\rangle \quad \Leftrightarrow \quad \hat{H}_\text{eff}(E)|\Psi_0\rangle = E|\Psi_0\rangle \\
    & |\Psi\rangle = \left(1+(E-H_{11})^{-1}T_{10})\right)|\Psi_0\rangle \\
\end{aligned}
\label{Psieff}
\end{equation}
Eq \ref{Heff} is used to construct the effective Hamiltonian in the reference space, and Eq \ref{Psieff} is used to evaluate the effective wave functions of reference states.

If $H_{11}$ is a very large matrix, the analytical expression of $(\omega-H_{11})^{-1}$ is prohibitively complicated. 
However, if the total Hamiltonian can be divided into the reference part and perturbative part, 
\begin{equation}
    H=H^0+V, \quad H^0= 
    \begin{pmatrix}
        H_{00}^0 & 0 \\
        0 & H_{11}^0
    \end{pmatrix}, \quad V = 
    \begin{pmatrix}
        V_{00} & V_{01} \\
        V_{10} & V_{11}
    \end{pmatrix}
\end{equation}
then the perturbative expansion of Eq \ref{Heff} with respect to $V$ can be obtained. 
Note that in low-order perturbations, $\omega$ is fixed as the unperturbed energy of reference states.
In high-order perturbations, the low-order correction to the reference states should be added to $\omega$. 
\begin{equation}
\begin{aligned}
    \omega & = \omega_0 + \omega_1 + \omega_2 + \omega_3 + ... \\
    \omega_n & = \left(\text{Energy of state interested in according to } H^{(n)}_\text{eff}\right) \sim O[V^n]
\end{aligned}
\end{equation}
For example, in our analysis, $\omega_i$ is the diagonal element of the states $|\uparrow\downarrow\rangle = \hat{d}^\dagger_{1\uparrow}\hat{d}^\dagger_{2\downarrow}\hat{p}^\dagger_\uparrow\hat{p}^\dagger_\downarrow|\mathrm{core}\rangle$ and $|\downarrow\uparrow\rangle = \hat{d}^\dagger_{1\downarrow}\hat{d}^\dagger_{2\uparrow}\hat{p}^\dagger_\uparrow\hat{p}^\dagger_\downarrow|\mathrm{core}\rangle$ in $H^{(i)}_\text{eff}$. 
Since $H^{(n)}_\text{eff} = V_{01}\left(\text{some matrix containing $\omega$}\right)V_{10} \sim V^n$, $H^{(n)}_\text{eff}$ contains at most the contribution of $\omega_{n-2} \sim V^{n-2}$. 
Therefore, the perturbation Hamiltonians $H_\text{eff}^{(n)}$ can be determined by a series of equations that do not contain self-references. 

\begin{equation}
\begin{aligned}
    H_\mathrm{eff}^{(0)} & = H_{00}, \quad H_\mathrm{eff}^{(1)} = V_{00} \\
    H_\mathrm{eff}^{(2)} & = V_{01}(\omega_0-H_{11}^0)^{-1}V_{10} \\
    H_\mathrm{eff}^{(3)} & = V_{01}(\omega_0-H_{11}^0)^{-1}(V_{11}-\omega_1)(\omega_0-H_{11}^0)^{-1}V_{10} \\
    H_\mathrm{eff}^{(4)} & = V_{01}(\omega_0-H_{11}^0)^{-1}(V_{11}-\omega_1)(\omega_0-H_{11}^0)^{-1}(V_{11}-\omega_1)(\omega_0-H_{11}^0)^{-1}V_{10} \\
    & + V_{01}(\omega_0-H_{11}^0)^{-1}(-\omega_2)(\omega_0-H_{11}^0)^{-1}V_{10} \\
    H_\mathrm{eff}^{(5)} & = V_{01}(\omega_0-H_{11}^0)^{-1}(V_{11}-\omega_1)(\omega_0-H_{11}^0)^{-1}(V_{11}-\omega_1)(\omega_0-H_{11}^0)^{-1}(V_{11}-\omega_1)(\omega_0-H_{11}^0)^{-1}V_{10} \\
    & + V_{01}(\omega_0-H_{11}^0)^{-1}(-\omega_2)(\omega_0-H_{11}^0)^{-1}(V_{11}-\omega_1)(\omega_0-H_{11}^0)^{-1}V_{10} \\
    & + V_{01}(\omega_0-H_{11}^0)^{-1}(V_{11}-\omega_1)(\omega_0-H_{11}^0)^{-1}(-\omega_2)(\omega_0-H_{11}^0)^{-1}V_{10} \\
    & + V_{01}(\omega_0-H_{11}^0)^{-1}(-\omega_3)(\omega_0-H_{11}^0)^{-1}V_{10}
\end{aligned}
\end{equation}
The general formula of $H_\text{eff}^{(n)}$ involves the strict integer composition $\mathcal{P}_\text{s}^{(n)}$ of $n\in \mathbb{N}$. 
\begin{equation}
\begin{aligned}
    \mathcal{P}_\text{s}^{(n,m)} & := \{(a_1,a_2,...,a_m)|a_i>0, a_i\in\mathbb{Z}, \forall i=1,...,m; \sum_{i=1}^m{a_i}=n\} \\
    \mathcal{P}_\text{s}^{(n)} & := \bigcup_{m=1}^n{\mathcal{P}_\text{s}^{(n,m)}}
\end{aligned}
\end{equation}
\begin{equation}
\begin{aligned}
    H_\text{eff}^{(n+2)} = \sum_{m=1}^{n}{\sum_{(a_1,...,a_m)\in\mathcal{P}_\text{s}^{(n,m)}}} & {V_{01}(\omega_0-H_{11}^0)^{-1}W_{a_1}(\omega_0-H_{11}^0)^{-1}W_{a_2} ...} \\
    & ...{W_{a_{m-1}}(\omega_0-H_{11}^0)^{-1}W_{a_m}(\omega_0-H_{11}^0)^{-1}V_{10}}, \quad n\in\mathbb{N}
\end{aligned}
\end{equation}
where $W_n$ is the $V^n$-order perturbative correction to $(\omega-H_{11})$.
\begin{equation}
    W_n=\begin{cases}
        V_{11} - \omega_1, & n = 1; \\
        -\omega_n, & \text{otherwise}. 
    \end{cases}
\end{equation}

In our calculations, there are some perturbative energy levels that are much higher than other states, with the energy spacing $U\rightarrow\infty$. 
Typical examples are the states with Cu 4f or 5d occupied, where the orbital energies of Cu 4f and 5d, denotes as $\epsilon_f$, are much larger than any states without Cu 4f or 5d occupation. 
In these cases, the energy spacing $U$ appear only in $H_{11}^0$. 
Therefore, at the $U\rightarrow\infty$ limit, the Laurent series of $G_{11}^0\equiv (\omega_0-H_{11}^0)^{-1}$ with respect to $U$ can be obtained. 
\begin{equation}
     G^0_{11} = G^0_{11,0} + G^0_{11,1}U^{-1} + G^0_{11,2}U^{-2} + ..., 
\end{equation}
Then the $U^{-m}$-order contribution to $H_\mathrm{eff}^{(n)}$, denoted as $H_\mathrm{eff}^{(n,m)}$, can be calculated as
\begin{equation}
    H_\mathrm{eff}^{(n+2,m)} = U^{-m}\sum_{p=1}^{n}\sum_{\substack{(a_1,...,a_p)\in\mathcal{P}_\text{s}^{(n,p)} \\ (b_1,...,b_{p+1})\in\mathcal{P}^{(m,p+1)}}}{V_{01}G^0_{11,b_1}W_{a_1}G^0_{11,b_2}W_{a_2}...W_{a_{p-1}}G^0_{11,a_p}W_{a_p}G^0_{11,a_{p+1}}V_{10}}
\end{equation}
where $\mathcal{P}^{(n,m)}$ denotes all the unstrict integer compositions of $n$ into $m$ parts, defined as
\begin{equation}
    \mathcal{P}^{(n,m)} := \{(a_1,a_2,...,a_m)|a_i\geq0, a_i\in\mathbb{Z}, \forall i=1,...,m; \sum_{i=1}^m{a_i} = n\}. 
\end{equation}

\section{Perturbation theory analysis on roles of C\MakeLowercase{u} 4\MakeLowercase{f} and 5\MakeLowercase{d}}\label{Sec3_PerturbAnalysis}

Cu 4f, 5d and 5p orbitals are found to contribute prominently to AFM coupling in our calculation. 
Cu 5d can be attributed to radial breathing of Cu d orbitals, which stems from the molecular integral
\begin{equation}
    (3d~3d|3d~5d) = \int{d\mathbf{r}_1d\mathbf{r}_2 \frac{\phi^*_{3d}(\mathbf{r}_1)\phi_{3d}(\mathbf{r}_1)\phi^*_{3d}(\mathbf{r}_2)\phi_{5d}(\mathbf{r}_2)}{|\mathbf{r}_1-\mathbf{r}_2|}}.
\end{equation}
However, the similar molecular integrals $(3d~3d|3d~4f)$ and $(3d~3d|3d~5p)$ are both zero, because 3d has even parity, while 4f and 5p both have odd parity.
Therefore, Cu 4f and 5p cannot be included in the orbital breathing framework, and a new theoretical explanation on their roles are required. 
From the following analysis we can see that, although all these orbitals reside on copper atoms, the mechanism they contributes to AFM coupling are quite different. 
Cu 4f are subject to strong on-site pair-hybridization due to a considerable pair-hopping and Hund coupling term $(3d~4f|3d~4f)$, while Cu 5p are more related to the radial and axial breathing of bridging O $p_\sigma$ orbital. 
A unified explanation via an effective model are presented as follows. 

We start from the $p$-$d$ model including Cu 3d$_{x^2-y^2}$ ($d$) and O 2p$_y$ ($p$), and add into the model two Cu high-energy orbitals denoted as $f$, which can be any 4f or 5p orbitals. 
For Cu 5p$_y$, considerable hopping with O 2p$_y$ is present, while for Cu 4f$_{x(x^2-y^2)}$, 4f$_{y(x^2-y^2)}$ and 4f$_{z(x^2-y^2)}$ orbitals, only the on-site exchange integrals with Cu 3d$_{x^2-y^2}$ are strong enough for consideration. 
The final model contains 5 orbitals with 4 electrons, and the molecular orbital (MO) integrals extracted from \textit{ab initio} Hamiltonian are used to construct the theoretical model. 
After discarding MO integrals with small absolute value ($<0.05 \mathrm{~E_h}$), the rest of \textit{ab initio} Hamiltonian encompasses the most important terms within these orbitals. 
The MO integrals with absolute value  are all discarded. 
\begin{enumerate}
    \item[(1)] Orbital energies, i.e. $\epsilon_d=(d|h|d)=0$, 
    $\epsilon_f=(f|h|f)=\begin{cases}
        2.432 \mathrm{~E_h}, & f=\text{Cu 5p}; \\
        5.167 \mathrm{~E_h}, & f=\text{Cu 4f}, 
    \end{cases}$
    and $\epsilon_p=(p|h|p)=-0.0887\mathrm{~E_h}$, where $\epsilon_f\gg\epsilon_d>\epsilon_p$. 
    Therefore, for the following analysis, the $\epsilon_f-\epsilon_d\rightarrow \infty$ limit is taken. 
    \item[(2)] On-site and nearest-neighbor Coulomb repulsion integrals, including $U_p=(pp|pp)=0.556\mathrm{~E_h}$, $U_{pd}=(dd|pp)=0.362\mathrm{~E_h}$ and $U_d=(dd|dd)=0.861\mathrm{~E_h}$. 
    These are associated with Coulomb repulsion terms $\hat{n}_{x\uparrow}\hat{n}_{x\downarrow}$ or $\hat{n}_x\hat{n}_y$. 
    The Coulomb repulsion involving $f$ orbitals, such as $U_f=(ff|ff)$ or $U_{df}=(dd|ff)$, are also large, but they do not participate in AFM coupling until $(\epsilon_f-\epsilon_d)^{-2}$ order, so they are neglected. 
    \item[(3)] Hopping integrals between d and p, which are related to hopping terms formulated as $\hat{h}_{dp}:=\sum_\sigma{(\hat{d}^\dagger_\sigma\hat{p}_\sigma + \hat{p}^\dagger_\sigma\hat{d}_\sigma)}$. 
    Due to the sensitive dependence of AFM coupling constant on model parameters, the hopping amplitudes are chosen according to \textit{ab initio} molecular integrals, which contain $t_{pd1}=(dp|dd)=0.1772\mathrm{~E_h}$, $t_{pd2}=(dp|pp)=0.0625\mathrm{~E_h}$ and $t_{pd3}=(d_1p|d_2d_2)=0.0537\mathrm{~E_h}$. 
    The simplest hopping term $t_{pd}=(d|h|p)$ is more common in literature (e.g. ref \citenum{plakida_high-temperature_2010}), but it is as small as $-0.012\mathrm{~E_h}$ according to \textit{ab initio} Hamiltonian. 
    In order to keep consistent with the model parameters related to f-levels which are also extracted from \textit{ab initio} integrals, we exploit a revised $p$-$d$ model which contains $t_{pd1}$, $t_{pd2}$ and $t_{pd3}$, which are associated with terms like $\hat{x}^\dagger\hat{h}_{pd}\hat{x} = \hat{h}_{pd}\hat{n}_x + \hat{n}_x\hat{h}_{pd} - (\delta_{xp}+\delta_{xd})\hat{h}_{pd}$.
    Although this model is more complicated in its formula, it is physically equivalent with the literature adopted version, and are connected with the \textit{ab initio} extracted values in a more straightforward way. 
    \item[(4)] Exchange integral between Cu 3d and O 2p, i.e. $K_{pd}=(dp|dp)=0.086\mathrm{~E_h}$, and the direct hopping between neighboring copper atoms, $t_{dd}=(d_1|h|d_2)=0.061\mathrm{~E_h}$.
    These integrals are related to $\hat{h}_{pd}^2 - \hat{n}_{d}-\hat{n}_p$, and $\hat{h}_{d1,d2}$ terms, respectively. 
    Both are crucial in the $J$ improvement due to Cu 5p. 
    \item[(5)] Hopping between O 2p and $f$, i.e. $t_{pf}=(f|h|p)$, which is important in Cu 5p$_y$ case ($0.044\mathrm{~E_h}$), but is negligible in Cu 4f cases. 
    The significance of f-p hopping is further enhanced if the radial breathing of O 2p onto 3p is considered, which increases the overlap and decrease the f-p energy level difference, and is beneficial to AFM coupling. 
    \item[(6)] Two-body exchange integral between Cu 3d and $f$, i.e. $K_{df}=(df|df)=0.119\mathrm{~E_h}$. This integral is associated with $\hat{h}_{fd}^2 - \hat{n}_{d}-\hat{n}_{f}$. 
    The pair-hopping in $K_{df}$ is the main driving force of Cu 4f contribution to AFM $J$. 
\end{enumerate}

Summarizing all the integrals above, the model Hamiltonian is formulated as $\hat{H} = \hat{H}_0 + \hat{H}_1$, as presented in the main text.
The unperturbed Hamiltonian $H_0$ includes the diagonal elements in SD basis, while the perturbation $H_1$ includes all the off-diagonal elements, corresponding to transitions between different configurations. 

\begin{table}[b]
\normalsize
\caption{\label{param_table}%
\normalsize
Low-energy effective parameters for different 4f or 5p orbitals. 
The 2-body parameters (e.g. $U$'s and $K$'s) are directly extracted from the 2-body bare \textit{ab initio} integrals. 
Other 4f or 5p components not shown all have very small $t_{pf}$ or $K_{df}$, and are neglected. 
The 1-body parameters (e.g. $\epsilon$'s and $t$'s) are extracted by calculating Fock matrix with core double occupation, i.e.
$(p|h|q)=h_{pq}+\sum_{r\in\{\mathrm{Occ}\}}{((pq|rr)-(pr|rq))}$, where $\{\mathrm{Occ}\}$ stands for occupied orbitals not included in the model. 
The unit is Hartree ($\mathrm{E_h}$). 
The 2-body integrals are written in chemical notation. $(pq|rs)= \int{d\mathbf{r}_1d\mathbf{r}_2p^*(\mathbf{r}_1)q(\mathbf{r}_1)\frac{1}{|\mathbf{r}_1-\mathbf{r}_2|}}r^*(\mathbf{r}_2)s(\mathbf{r}_2). $
%
}
\begin{ruledtabular}
{\renewcommand{\arraystretch}{1.4}
\begin{tabular}{cdddd}
\textrm{Parameter name} &
\multicolumn{1}{c}{Cu 4f$_{x(x^2-y^2)}$} &
\multicolumn{1}{c}{Cu 4f$_{z(x^2-y^2)}$} &
\multicolumn{1}{c}{Cu 4f$_{y(x^2-y^2)}$} &
\multicolumn{1}{c}{Cu 5p$_y$}\\
\colrule
$\epsilon_p-\epsilon_d$ & -0.089 & -0.089 & -0.089 & -0.089 \\
$\epsilon_f-\epsilon_d$ & 5.167 & 5.224 & 5.168 & 2.431 \\
$U_d=(dd|dd)$ & 0.861 & 0.861 & 0.861 & 0.861 \\
$U_{pd}=(dd|pp)$ & 0.362 & 0.362 & 0.362 & 0.362 \\
$U_{dd}=(d_1d_1|d_2d_2)$ & 0.168 & 0.168 & 0.168 & 0.168 \\
$U_p=(pp|pp)$ & 0.556 & 0.556 & 0.556 & 0.556 \\
$t_{pd1}=(dp|dd)$ & 0.177 & 0.177 & 0.177 & 0.177 \\
$t_{pd2}=(dp|pp)$ & 0.0625 & 0.0625 & 0.0625 & 0.0625 \\
$t_{pd3}=(d_1p|d_2d_2)$ & 0.0537 & 0.0537 & 0.0537 & 0.0537 \\
$K_{pd}=(dp|dp)$ & 0.0861 & 0.0861 & 0.0861 & 0.0861 \\
$t_{dd}=(d_1|d_2)$ & 0.0608 & 0.0608 & 0.0608 & 0.0608 \\
$t_{pf}=(f|h|p)$ & -0.009 & -0.002 & 0.000 & 0.043 \\
$K_{df}=(df|df)$ & 0.119 & 0.093 & 0.119 & 0.027
\end{tabular}
}
\end{ruledtabular}
\end{table}

The parameter values are listed in Table \ref{param_table}. 
$t_{pf}$ and $K_{df}$ are the only parameters related to $f$ orbitals, which are responsible for Cu 4f and 5p, respectively. 
The Hilbert space contains (4e,5o), and perturbation analysis can be performed with the downfolding technique described in Sec \ref{Sec2_Downfold}. 
To calculate the downfolding effect on $J$, one can select the reference states as $|d_{1\uparrow}d_{2\downarrow}p^2\rangle = \hat{d}^\dagger_{1\uparrow}\hat{d}^\dagger_{2\downarrow}\hat{p}^\dagger_\uparrow\hat{p}^\dagger_\downarrow|\mathrm{vac}\rangle$ and $|d_{1\downarrow}d_{2\uparrow}p^2\rangle = \hat{d}^\dagger_{1\downarrow}\hat{d}^\dagger_{2\uparrow}\hat{p}^\dagger_\uparrow\hat{p}^\dagger_\downarrow|\mathrm{vac}\rangle$, where the energy difference between the open-shell singlet and triplet is AFM $J$. 
All hopping and exchange terms in $\hat{H}_1$ are viewed as perturbations.

The perturbation calculation of $\hat{H}_1$ yield the lowest-order term of AFM coupling as the $(H_1)^2$ order term. 
\begin{equation}
    J^{(2)} = \frac{4t_{dd}^2}{U_\text{eff}}, 
    \qquad U_\text{eff} = U_d-U_{dd} = E[d_1^2p^2] - E[d_{1\uparrow}d_{2\downarrow}p^2]
\end{equation}
Here we use $J^{(m)}$ to denote the $(H_1)^m$ order term of $J$. 
The leading contribution of $K_{df}$ appears in the $(H_1)^4(\epsilon_f-\epsilon_d)^{-1}$-order term, which dominates the Cu 4f contribution to AFM coupling. 
The leading contribution of $t_{pf}$, however, appears in $(H_1)^4(\epsilon_f-\epsilon_d)^{-2}$ and $(H_1)^5(\epsilon_f-\epsilon_d)^{-1}$-order term, which is responsible for Cu 5p effect. 
\begin{equation}
\begin{aligned}
    J^{(4,1)}_K & = \frac{2K_{df}^2t_{dd}^2}{(\epsilon_f-\epsilon_d)U_\text{eff}^2} \\
    J^{(4,2)}_{t} & = \frac{16t_{pf}^2t_{dd}^2}{(\epsilon_f-\epsilon_d)^2U_\text{eff}} \\
    J^{(5,1)}_t & = \frac{16t_{pf}^2t_{2}}{(\epsilon_f-\epsilon_d)U_\mathrm{1CT}^3}\left(\frac{t_{dd}t_{1}U_\mathrm{1CT}}{U_\mathrm{eff}} + t_{2}(K_{pd}-t_{dd})\right) \\
    & t_{1} = \langle d_{1\downarrow}d_2^2p_\uparrow|\hat{H}|d_2^2p^2\rangle = 2t_{pd3} + t_{pd2} \\
    & t_{2} = \langle d_{1\downarrow}d_{2\uparrow}p^2|\hat{H} |d_{1\downarrow}d_2^2p_\uparrow\rangle = t_{pd1} - t_{pd3} - t_{pd2}
\end{aligned}
\end{equation}
where $J^{(m,n)}$ denotes $(H_1)^m(\epsilon_f-\epsilon_d)^{-n}$ order perturbation, and the subscript $K$ and $t$ denote $K_{df}$-related term and $t_{pf}$-related term, respectively. 
From the parameters (Table \ref{param_table}) one can verify that all of these are of positive sign. 
Therefore, the correlation of Cu 4f and 5p leads to the enhancement of $J$ through different mechanisms, which involve the direct exchange $K_{df}$ and direct hopping $t_{pf}$, respectively.

It is worth noting that the leading contribution of $t_{pf}$ stems from $t_{dd}$ and $K_{pd}$, which are not present in $p$-$d$-models that is usually adopted in literature, e.g. ref \citenum{plakida_high-temperature_2010}. 
Actually, we can observe an oversimplified model where $t_{dd}$ and $K_{pd}$ are neglected can preserve the superexchange enhancement due to Cu 4f, but cannot capture the contribution from Cu 5p. 
If $t_{dd}$, $K_{pd}$ and $dp$-hopping terms other than $t=t_{pd1}$ are discarded, the leading term of AFM coupling is the $t^4$ term. 
\begin{equation}
\begin{aligned}
    J^{(4)} & = \frac{8t^4}{U_\mathrm{1CT}^2U_\mathrm{2CT}} \\
    U_{\mathrm{1CT}} & = \epsilon_d-\epsilon_p+U_d-U_{pd}+U_{dd}+U_p = E[d_1^2d_{2\downarrow}p_{\uparrow}] -E[d_{1\uparrow}d_{2\downarrow}p^2] \\
    U_{\mathrm{2CT}} & = 2\epsilon_d-2\epsilon_p+2U_d-4U_{pd}+3U_{dd}-U_p = E[d_1^2d_2^2] - E[d_{1\uparrow}d_{2\downarrow}p^2] \\
\end{aligned}
\end{equation}
In this model, the contributive effect of Cu 4f orbitals can be explained in the $V^6(\epsilon_f-\epsilon_d)^{-1}$ term. 
\begin{equation}
    J^{(6,1)}_K = \frac{8K_{df}^2t^4(U_\mathrm{2CT}+U_\mathrm{1CT})}{(\epsilon_f-\epsilon_d)U_\mathrm{2CT}^2U_\mathrm{1CT}^3}
\end{equation}
However, this model does not contain the contributive effect of $t_{pf}$, which occurs in the \textit{ab initio} calculation benchmarks, and necessarily relies on $t_{dd}$ and $K_{pd}$. 
Therefore, the more complicated model formulated in this work is required for a complete explanation on Cu 4f and 5p contribution.

\section{FCIQMC results}\label{SecIV_FCIQMC}

FCIQMC is a state-of-the-art electronic method that accurately solves the FCI wavefunction in a non-perturbative way. 
In this work, FCIQMC calculations are conducted on several CAS spaces where NEVPT2 calculations are performed, with the CAS and PT orbitals both included in FCIQMC orbital space. 
In this way, the perturbative analysis on PT space in NEVPT2 method can be confirmed using FCIQMC. 
The results are listed in Table \ref{FCIQMC_result}. 
All FCIQMC runs reach convergence with respect to walker number within $1\text{meV}$. 

\begin{table}[h!]
\normalsize
\caption{\label{FCIQMC_result}%
\normalsize
FCIQMC results of several CAS. 
``$J$ ($N_w=x$)'' means the AFM $J$ calculated with the number of walkers fixed to $x$ (M=$10^6$). 
CAS-DMRG includes Cu 3d4d and O$_b$ 2p3p orbitals.
The unit of $J$ is meV. 
%
%
}
\begin{ruledtabular}
{\renewcommand{\arraystretch}{1.1}
\begin{tabular}{p{40mm}<{\centering}ccccc}
\textrm{CAS description} &
\textrm{CAS size} &
$J$ (1M) &
$J$ (3M) &
$J$ (10M) &
$J$ (30M) \\
\colrule
CAS-DMRG & (24e,26o) & $95.7\pm2.0$ & $94.1\pm1.3$ & $92.9\pm0.8$ & \textemdash \\
CAS-DMRG La 5d & (24e,46o) & $107.3\pm3.2$ & $96.0\pm2.1$ & $94.9\pm1.3$ & $95.5\pm0.4$ \\
CAS-DMRG Cu 4f5p5d & (24e,56o) & \textemdash & \textemdash & $146.8\pm1.4$ & $146.6\pm1.0$ \\
\end{tabular}
}
\end{ruledtabular}
\end{table}

The AFM coupling calculated by FCIQMC within CAS(24e,26o), which contains Cu 3d, 4d and O 2p, 3p, is 92.9(8) meV.
This verifies the negligible error of DMRG wavefunction, and validates DMRG wavefunction as the reference for multi-reference perturbation calculations. 
The FCIQMC $J$ in CAS's that contain all important atomic shells (Cu 4f 5p 5d), or an unimportant atomic shell (La 5d), are also obtained. 
It turns out that the contribution to $J$ of La 5d is still negligible in FCIQMC calculation. 
The contribution of Cu 4f, 5p and 5d, is also significant ($\sim 50\text{ meV}$), which is slightly larger than the contribution from NEVPT2 ($\sim35 \text{ meV}$). 
Therefore, NEVPT2 is capable of qualitatively grasping the physically important atomic shells for AFM coupling. 

\section{Cluster structure and basis sets}\label{SecV_structure}

Basis set and ECP: 

Cu1: cc-pVDZ

O1: cc-pVDZ

Cu2: CRENBS with ECP

O2: cc-pVDZ

O3: cc-pVDZ

O4: cc-pVDZ

O5: cc-pVDZ@3s2p

La1: ECP46MWB@2s2p1d with ECP

La2: CRENBS with ECP

\clearpage
Cluster structure: 
\begin{verbatim}
O1     -0.000000000     0.000000000     0.092050000
Cu1     0.006363961     1.904945669     0.000000000
Cu1    -0.006363961    -1.904945669     0.000000000
La1     1.878187334    -0.020394374    -1.814700000
La1    -1.878187334     0.020394374    -1.814700000
La1     1.931704003     0.033122296     1.814700000
La1    -1.931704003    -0.033122296     1.814700000
O3      1.898581707    -1.898581707     0.092050000
O3     -1.898581707     1.898581707     0.092050000
O3      1.911309630     1.911309630    -0.092050000
O3     -1.911309630    -1.911309630    -0.092050000
O2      0.112137236    -1.786444471    -2.459050000
O2     -0.112137236     1.786444471    -2.459050000
O2     -0.124865158    -2.023446866     2.459050000
O2      0.124865158     2.023446866     2.459050000
O4      3.809891337     0.012727922    -0.092050000
O3      0.012727922     3.809891337    -0.092050000
O4     -3.809891337    -0.012727922    -0.092050000
O3     -0.012727922    -3.809891337    -0.092050000
Cu2    -3.803527376     1.892217746     0.000000000
Cu2     3.803527376    -1.892217746     0.000000000
Cu2    -3.816255298    -1.917673591     0.000000000
Cu2     3.816255298     1.917673591     0.000000000
La2     1.890915256     3.789496963     1.814700000
La2    -1.890915256    -3.789496963     1.814700000
La2    -1.918976081     3.776769041    -1.814700000
La2     1.918976081    -3.776769041    -1.814700000
La2     1.865459412    -3.830285711     1.814700000
La2    -1.865459412     3.830285711     1.814700000
La2    -1.944431925    -3.843013633    -1.814700000
La2     1.944431925     3.843013633    -1.814700000
La2     0.020394374    -1.878187334     4.760300000
La2    -0.020394374     1.878187334     4.760300000
La2     0.033122296     1.931704003    -4.760300000
La2    -0.033122296    -1.931704003    -4.760300000
O5      3.797163415    -3.797163415     0.092050000
O5     -3.797163415     3.797163415     0.092050000
O5      3.822619259     3.822619259     0.092050000
O5     -3.822619259    -3.822619259     0.092050000
Cu2     0.019091883     5.714837006     0.000000000
Cu2    -0.019091883    -5.714837006     0.000000000
O5      1.885853785    -5.708473045    -0.092050000
O5     -1.885853785     5.708473045    -0.092050000
O5      5.708473045    -1.885853785    -0.092050000
O5     -5.708473045     1.885853785    -0.092050000
O5     -5.721200967    -1.924037552     0.092050000
O5     -1.924037552    -5.721200967     0.092050000
O5      1.924037552     5.721200967     0.092050000
O5      5.721200967     1.924037552     0.092050000
O5      0.025455844     7.619782674     0.092050000
O5     -0.025455844    -7.619782674     0.092050000
\end{verbatim}

\ifmain

\else

\bibliography{main.bib}

\end{document}

\fi
%

\end{document}